\documentclass[fleqn,usenatbib]{mnras}

\usepackage{newtxtext,newtxmath}

\usepackage[T1]{fontenc}
\usepackage{ae,aecompl}

\usepackage{graphicx}	
\usepackage{amsmath}	
\usepackage{hyperref}
\usepackage{chngcntr}
\counterwithout{figure}{section}
\counterwithout{table}{section}
\usepackage{nccmath}
\usepackage{placeins}
\usepackage{adjustbox}
\usepackage{array}

\title[Effect of running on large-scale structure]{The BAHAMAS project: Effects of a running scalar spectral index on large-scale structure}

\author[S. G. Stafford et al.]{
Sam G. Stafford$^{1}$\thanks{E-mail: S.Stafford@2014.ljmu.ac.uk},
Ian G. McCarthy$^{1}$\thanks{E-mail: i.g.mccarthy@ljmu.ac.uk},
Robert A. Crain$^{1}$, Jaime Salcido$^{1}$, \newauthor Joop Schaye$^{2}$, Andreea S. Font$^{1}$, Juliana Kwan$^{1}$, Simon Pfeifer$^{1}$
\\
$^{1}$Astrophysics Research Institute, Liverpool John Moores University, 146 Brownlow Hill, Liverpool L3 5RF, UK \\
$^{2}$Leiden Observatory, Leiden University, P. O. Box 9513, 2300 RA Leiden, The Netherlands
}

\date{Accepted 2019 December 24. Received 2019 November 15; in original form 2019 July 22}

\pubyear{2019}

\begin{document}
\label{firstpage}
\pagerange{\pageref{firstpage}--\pageref{lastpage}}
\maketitle

\begin{abstract}
Recent analyses of the cosmic microwave background (CMB) and the Lyman-$\alpha$ forest indicate a mild preference for a deviation from a power law primordial matter power spectrum (a so-called `running').
We introduce an extension to the \texttt{BAHAMAS} suite of simulations to explore the effects that a running scalar spectral index has on large-scale structure (LSS), using $Planck$ CMB constraints to initialize the simulations. 
We focus on 5 key statistics: i) the non-linear matter power spectrum ii) the halo mass function; iii) the halo two-point auto correlation function; iv) total mass halo density profiles; and v) the halo concentration-mass relation. We find that the matter power spectrum in a $Planck$-constrained running cosmology is affected on all $k-$scales examined in this study. 
These effects on the matter power spectrum should be detectable with upcoming surveys such as LSST and Euclid. 
A positive running cosmology leads to an increase in the mass of galaxy groups and clusters, with the favoured negative running leading to a decrease in mass of lower-mass ($M \lesssim 10^{13} \textrm{M}_{\odot}$) haloes, but an increase for the most massive ($M \gtrsim 10^{13} \textrm{M}_{\odot}$) haloes. Changes in the mass are generally confined to $5$-$10\%$ which, while not insignificant, cannot by itself reconcile the claimed tension between the primary CMB and cluster number counts.We also demonstrate that the observed effects on LSS due to a running scalar spectral index are separable from those of baryonic effects to typically a few percent precision. 
\end{abstract}

\begin{keywords}
cosmology: large-scale structure of Universe -- cosmology: cosmological parameters -- cosmology: inflation
\end{keywords}



\section{Introduction}
The standard model of cosmology is remarkably successful at describing how structure in the Universe formed and, with the recent $Planck$ mission, the model has been validated and constrained to an unprecedented precision (see \citealt{PlanckXIII}). 
One of the remarkable aspects of this model, termed the $\Lambda$CDM model, is that it can be described in full by just 6 independent adjustable parameters. 
However, with the wealth of observational data available today, being taken with ever more precise instruments, a few interesting tensions have arisen with some of the derived parameters of this model. 
For example, there is the well-known tension in the Hubble constant ($H_0$) with local measurements \citep[e.g.][]{Bonvin2017, Riess2018}, preferring a higher value for $H_0$ than the value obtained via the analysis of the primary Cosmic Microwave Background (CMB) and Baryon Acoustic Oscillations (BAO) (e.g. \citealt{PlanckXIII}). 
There also exists a mild tension when comparing various large-scale structure (LSS) joint constraints on the matter density $\Omega_{\textrm{m}}$ and $\sigma_8$ (the linearly evolved amplitude of density perturbations on 8 Mpc $h^{-1}$ scales) to the constraints on these quantities from $Planck$ measurements of the CMB. 
In particular, there are a number of LSS data sets which appear to favour relatively low values for $\Omega_{\textrm{m}}$ and/or $\sigma_8$ \citep[see e.g.][]{Heymans2013, PlanckXXIV, Hildebrandt2017, Joudaki2018, McCarthy2018, Abbott2019}. 
In addition to these tensions with low-redshift probes, a number of studies have demonstrated that there are a few mild internal tensions in the $Planck$ data/modelling itself \citep[see e.g.][]{Addison2016,Planck2017LI}. 
Together, these tensions, if they are not just statistical fluctuations or unaccounted for systematic errors in the analyses, could be signs of interesting new physics. 

\begin{figure}
    \centering
    \includegraphics[width=\columnwidth]{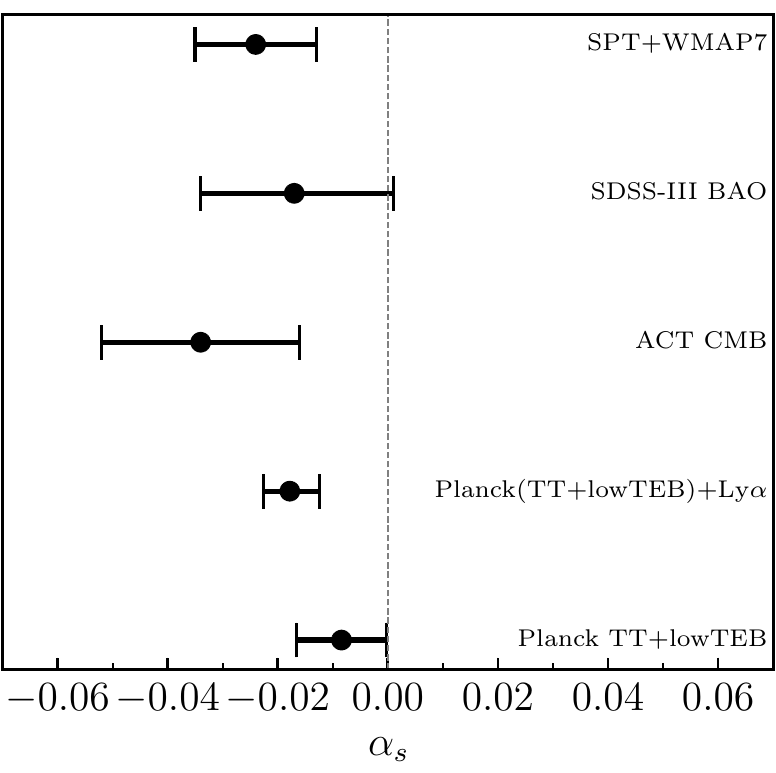}
    \vspace{-0.5cm}
    \caption{Summary of the observational constraints discussed in the introduction to this study. The vertical dashed line indicates no running. There is currently a mild preference for a negative value for $\alpha_s$.}
    \label{fig:current_constraints}
\end{figure}

There are many possible ways to extend the standard model of cosmology that could potentially reconcile some of the above tensions, including (non-minimal) massive neutrino cosmologies \citep[e.g.][]{Battye2014, Beutler2014, Wyman2014, McCarthy2018}, dynamical dark energy models \citep[e.g.][]{DiValentino2017, Yang2019}, and deviations from general relativity (`modified gravity') \citep[e.g.][]{DeFelice2010, Li2012, Nunes2018} to name just a few. One of the possible extensions that is perhaps less commonly discussed is a running of the scalar spectral index of the primordial power spectrum.
The primordial spectral index, $n_s$, of scalar perturbations is most often assumed to be independent of scale (i.e., that the primordial power spectrum is a pure power law). 
However, virtually all models of inflation predict at least some small level of scale dependence in $n_s$. For example, the simplest models of inflation (e.g., single-field, slow-roll inflation) predict a running of $n_s$ of order $\mathcal{O}(1 - n_s)^2$ (\citealt{Kosowsky1995}) (this would be $\sim$ 0.001 for $Planck$ CMB constraints on $n_s$), where the scale dependence of the spectral index is given by $n_s(k) = n_s(k_0) + \alpha_s\ln(k/k_0)$, with $\alpha_s$ being the `running' of the scalar spectral index and $k_0$ the pivot scale.

Using the $Planck$ 2015 full mission temperature data, $\alpha_s$ is constrained to have a central value of: $\alpha_s = -0.00841^{+0.0082}_{-0.0082}$ (68$\%$ CL $Planck TT+lowTEB$). 
That is, there is a very mild preference for a negative running from the $Planck$ CMB data. 
\citet{Palanque-Delabrouille2015} combine their measurements of the Lyman-alpha forest with the $Planck$ 2015 full mission temperature data and low multipole polarisation data to find a much stronger preference for a negative running: $\alpha_s = -0.0178^{+0.0054}_{-0.0048}$ (68$\%$ CL $ Planck (TT + lowTEB) + Ly\alpha$). 
They do note, however, that this result may be due to unaccounted for systematics in their measurements. 
In addition to this, there have been other independent measurements (i.e., measurements which have not used $Planck$ CMB data) of a running scalar spectral index, at varying levels of significance e.g.: $\alpha_s = -0.034^{+0.018}_{-0.018}$ ($ACT$ $CMB$) (\citealt{Dunkley2011}); $\alpha_s = -0.017^{+0.018}_{-0.017}$ ($SDSS-III$ $BAO$)(\citealt{Zhao2013}); $\alpha_s = -0.024^{+0.011}_{-0.011}$ ($SPT+WMAP7$)(\citealt{Hou2014}). A visual summary of the current empirical constrains on the running is shown in Fig. \ref{fig:current_constraints}. 

Note that, since different models of inflation predict different scale dependencies, measurements of $\alpha_s$ and its running (i.e., running of the running) can be used to constrain, or possibly rule out, models of inflation (see \citealt{Escudero2016} for a discussion on the impact on inflationary models in the light of the $Planck$ results).

Given these results, and the possible tensions which currently exist, it is interesting to see what effects a running spectral index that is within observational constraints would have on the LSS that we see in the Universe today. 
This is the aim of this present study. 
We use direct numerical simulations which allow us to accurately model the non-linear growth of structure for cosmologies with a running spectral index imprinted into the initial conditions (ICs) to explore the differences which arise compared to the standard $\Lambda$CDM model.
We do this using a new extension to the \texttt{BAHAMAS} suite of cosmological hydrodynamic simulations \citep{McCarthy2017,McCarthy2018}, which is described below.
In this study, we will explore 5 main statistics to investigate the effects due to a $\Lambda \alpha_s$CDM model. These statistics are: the non-linear total matter power spectrum, the halo mass function (HMF), the halo two-point autocorrelation function, total halo mass density profiles, and the halo concentration-mass relation. 
We will also examine how separable the effects of a running scalar spectral index are from the effects due to baryonic physics, to assess if they can be treated as two separate multiplicative corrections to the standard model.

This paper is organized as follows: in Section \ref{sec:Sims} we present a brief summary of the simulations used, as well as our parameter selection method. 
In Section \ref{sec:LSS} we examine the effects of a running scalar spectral index on the LSS present in the simulations, including the HMF, the two-point autocorrelation function, and the total matter power spectrum. 
In Section \ref{sec:Internal_effects} we show the effects a running scalar spectral index has on certain internal halo properties, such as the total mass density profiles, and the concentration-mass relation. 
In Section \ref{sec:separability:checks} we present a separability test of the effects due to a running scalar spectral index, and baryonic physics, to assess if these two processes can be treated independently. 
In Section \ref{sec:conclusions} we summarise and discuss our findings.

\section{Simulations}
\label{sec:Sims}
\subsection{\texttt{BAHAMAS}}
\label{BAHAMAS}
This study extends the current \texttt{BAHAMAS} suite of cosmological hydrodynamic simulations. In this extension, a running scalar spectral index is incorporated into the initial conditions.

The \texttt{BAHAMAS} suite of cosmological simulations described in \cite{McCarthy2017} (see also \citealt{McCarthy2018}) consists of 400 comoving Mpc $h^{-1}$ on a side, periodic box, smooth particle hydrodynamics (SPH) simulations containing 2 $\times$ 1024$^3$ particles.
The present study adds to the pre-existing suite of \texttt{BAHAMAS} simulations with a new subset, whose initial conditions are based on the \textit{Planck} maximum-likelihood cosmological parameters derived from the \textit{Planck} 2015 data release \citep{PlanckXIII}.
The cosmological parameters of each run were varied, including values for the running of the scalar spectral index $\alpha_s$ (the method for how the cosmologies were chosen is discussed in detail in Section \ref{ParamSelection}). 
The Boltzmann code \texttt{CAMB}\footnote{\href{http://camb.info/}{http://camb.info/}} (\citealt{Lewis2000}, August 2018 version) was used to compute the transfer functions, and a modified version of N-GenIC was to used to create the initial conditions for the simulations, which start at a redshift of $z =$ 127. N-GenIC has been modified to include second-order Lagrangian Perturbation Theory corrections alongside support for massive neutrinos\footnote{\href{https://github.com/sbird/S-GenIC}{https://github.com/sbird/S-GenIC}}. 
Note that when producing the initial conditions, we use the separate transfer functions computed by \texttt{CAMB} for each individual component (i.e., baryons, neutrinos, and CDM) for the hydrodynamical simulations.  
Note also that, when producing the initial conditions for each of the simulations, the same random phases are used for each, implying that any comparisons made between the different runs are not subject to cosmic variance complications.

The simulations were carried out using a modified version of the Lagrangian TreePM-SPH code \texttt{GADGET3} (last described in \citealt{Springel2005}). This is a Lagrangian code used to calculate the gravitational and hydrodynamic forces on a system of particles. It was modified to include new subgrid physics as part of the OWLS project (see section 3 of \citealt{Schaye2010}).
The gravitational softening is fixed to 4 kpc $h^{-1}$ (in physical coordinates for $z \leq 3$ and in comoving coordinates at higher redshifts) and the SPH smoothing is done using the nearest 48 neighbours. The \texttt{BAHAMAS} run used here for a $Planck$ maximum-likelihood cosmology (with no running of the scalar spectral index) has dark matter and (initial) baryon particle mass of $\approx$ 4.36 $\times$ 10$^{9}$ $\textrm{M}_{\odot}$ $h^{-1}$ and $\approx$ 8.11 $\times$ 10$^{8}$ M$_{\odot}$ $h^{-1}$ respectively.
The particle masses for the other cosmologies in this suite do not differ much from these values but can be found in Table \ref{table:part_masses} (note that these slight differences in particle mass are due to the slight differences in $\Omega_m$ and $h$).

This suite of \texttt{BAHAMAS} simulations also uses a massive neutrino extension, described in \cite{McCarthy2018}. 
Here the simulations incorporate the {\it minimum} summed neutrino mass equal to $\Sigma M_{\nu} = 0.06$ eV implied by the results of atmospheric and solar oscillation experiments when adopting a normal hierarchy of masses \citep[][]{LesgourguesJ.Pastor2006}. We adopt the minimum neutrino mass for consistency, as this was what was adopted in the $Planck$ analysis when constraining the running of the scalar spectral index.
To model the effects of massive neutrinos on both the background expansion rate and the growth of density fluctuations, the semi-linear algorithm developed by \cite{Ali-Ha2013} \citep[see also][]{Bond1980, Ma1995, Brandbyge2008, Brandbyge2009, Bird2012} was implemented in the \texttt{GADGET3} code. This algorithm computes neutrino perturbations on the fly at each time step (see \citealt{McCarthy2018} for further details).
A study into the combined and separate effects of neutrino free-streaming and baryonic physics on collapsed haloes within \texttt{BAHAMAS} can be found in \citet{Mummery2017}.
In addition to neutrinos, all of the \texttt{BAHAMAS} runs (with or without massive neutrinos, or a running scalar spectral index) also include the effects of radiation when computing the background expansion rate. 

Note that for each hydro simulation, we also produce a corresponding `dark matter-only' simulation, where the collisionless particles follow the same total transfer function as used in the hydro simulations\footnote{As shown in \citet{VanDaalen2020} (see also \citealt{Valkenburg2017}), this setup leads to a small $\sim1\%$ offset in the amplitude of the $z=0$ matter power spectrum of the hydro simulations with respect to the dark matter only counterpart. This offset can be removed by instead using a dark matter-only simulation with two separate fluids (one with the CDM transfer function and the other with the baryon transfer function). This is unnecessary for the purposes of the present study, as we are only interested in the {\it relative} effects of different values of the running on LSS.}. These have the same cosmologies and initial phases as the hydro runs, but a dark matter particle mass of $\approx$ 5.17 $\times$ 10$^{9}$ M$_{\odot}h^{-1}$ (a complete list of dark matter particle masses for these simulations can also be found in Table \ref{table:part_masses}).

\subsection{Baryonic physics}
As in the original \texttt{BAHAMAS} suite of simulations, this extension also includes prescriptions for various `subgrid' processes, including: metal-dependent radiative cooling (\citealt{Wiersma2009a}); star formation (\citealt{Schaye2008}) and stellar evolution; mass loss and chemical enrichment from Type II and Ia supernovae, Asymptotic Giant Branch (AGB) and massive stars (\citealt{Wiersma2009}). Furthermore, the simulations include prescriptions for stellar feedback (\citealt{DallaVecchia2008}) and supermassive black hole growth and AGN feedback (\citealt{Booth2009}, which is a modified version of the model originally developed by \citealt{Springel2005b}).

As explained by \cite{McCarthy2017} and discussed in Section \ref{sec:separability:checks}, \texttt{BAHAMAS} is calibrated to reproduce the present-day galaxy stellar mass function for $M_* > 10^{10}M_{\odot}$ and the amplitude of the gas mass fraction-halo mass relation of groups and clusters, as inferred from high-resolution X-ray observations (note that synthetic X-ray observations of the original simulations were used to make a like-with-like comparison). The latter is particularly important for large-scale structure, since hot gas dominates the baryon budget of galaxy groups and clusters. To match these observables, the feedback parameters which control the efficiencies of the stellar and AGN feedback were adjusted. We have verified that the changes in cosmology explored here do not affect the calibration of the simulations, as such we have left these parameters at their calibrated values from \cite{McCarthy2017} for the present study.

\subsection{Running of the scalar spectral index}
As mentioned in the introduction, this study looks into an extension to the standard model of cosmology in the form of adding a scale dependence to the spectral index $n_s$ of the primordial matter power spectrum. This results in a modification to the equation for the primordial matter power spectrum.
\begin{ceqn}
\begin{align}
    \label{eq:Pk}
    P_s(k) = A_s(k_0)\left(\frac{k}{k_0}\right)^{n_s(k_0)+\alpha_s'(k)},
\end{align}
\end{ceqn}
\noindent where $\alpha_s'(k)\equiv (\alpha_s/2)\ln(k/k_0)$, $\alpha_s$ is the running of the scalar spectral index, which is defined as d$n_s$/d$\ln(k)$. The pivot scale $k_0$ is the scale at which the amplitude of the power spectrum ($A_s$) and the spectral index ($n_s$) are defined. In this study we adopt the same pivot scale as was used for the cosmological parameter estimation of \cite{PlanckXIII}: $k_0$ = 0.05 Mpc$^{-1}$ . 

\begin{figure}
    \centering
    \includegraphics[width=\columnwidth]{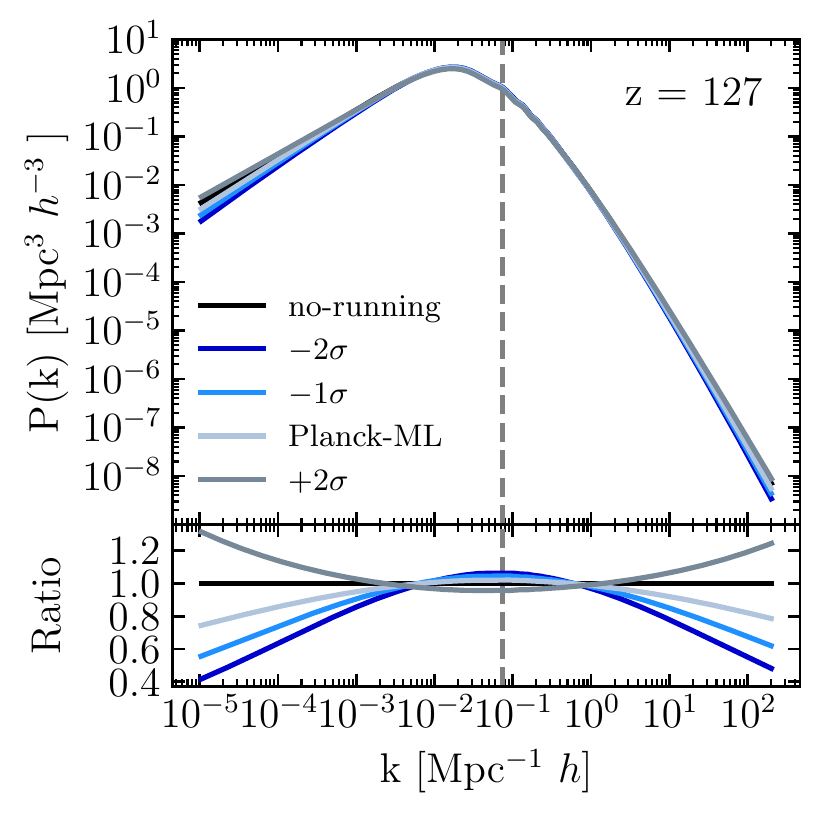}
    \vspace{-0.5cm}
    \caption{Linear matter power spectrum at $z$ = 127, as computed by \texttt{CAMB} for the 5 models presented in Table \ref{table:cosmoparams}. The lines here are coloured by the adopted value of $\alpha_s$. The vertical dashed grey line indicates the pivot scale $k_0$. As expected, the largest effect on the power spectrum is on the largest and smallest scales. At these scales, a negative running leads to less power compared with a no-running cosmology, whereas the positive running cosmology predicts more power on these scales.}
    \label{fig:Pk}
\end{figure}

The effect that the addition of this parameter has on the linear matter power spectrum can be computed using \texttt{CAMB} and the result is shown in Fig. \ref{fig:Pk}. This plot shows directly how large the effects can be on both small and large scales, with the negative (positive) running cosmologies showing a suppression (enhancement) of power on these scales. In terms of overdensities, this can be thought of as a smoothing out (amplification) of these overdensities for a negative (positive) running. 
There is, however, an interesting region near the pivot scale, where these effects are actually reversed, and a negative running leads to an amplification of power and a positive running leads to a suppression.
This behaviour is due to how the other cosmological parameter values are chosen, as described in Section \ref{ParamSelection}. 

Note that $\alpha_s$ is not the only additional parameter predicted by inflation.  There is also the so-called running of the running, $\beta$. This parameter adds a second-order scale-dependence to the spectral index in the form of: $\beta_s \equiv d^2n_s/d\ln(k)^2$, and leads to a power spectrum of the form:
\begin{ceqn}
    \begin{align}
        P_s(k) = A_s(k_0)\left(\frac{k}{k_0}\right)^{n_s(k_0)+\alpha_s'(k) + \beta_s'(k)},
    \end{align}
\end{ceqn}
\noindent where $\beta_s' \equiv (\beta_s/6)(\ln(k/k_0))^2$ and all other terms are as previously defined. Similarly to $\alpha_s$, analysis of the $Planck$ 2015 data indicates a mild preference for a non-zero running of the running, with $\beta_s = 0.029^{+0.015}_{-0.016}$ (68$\%$ CL $Planck TT+lowTEB$) (\citealt{PlanckXX}). In the present study we focus just on the first-order effect and we leave an exploration of running of the running on large-scale structure for future work.

\subsection{Cosmological parameter selection}
\label{ParamSelection}
To generate a set of cosmological parameters for the simulations, this study makes use of the publicly-available set of \cite{PlanckXIII} Markov chains, in particular those which include $\alpha_s$ as a free parameter. 
The parameter chains were produced using CosmoMC\footnote{\href{https://cosmologist.info/cosmomc/}{https://cosmologist.info/cosmomc/}} using a fast-slow dragging algorithm (\citealt{Neal2005}) and have already had burn-in removed. 

\begin{figure}
    \centering
    \includegraphics[width=\columnwidth]{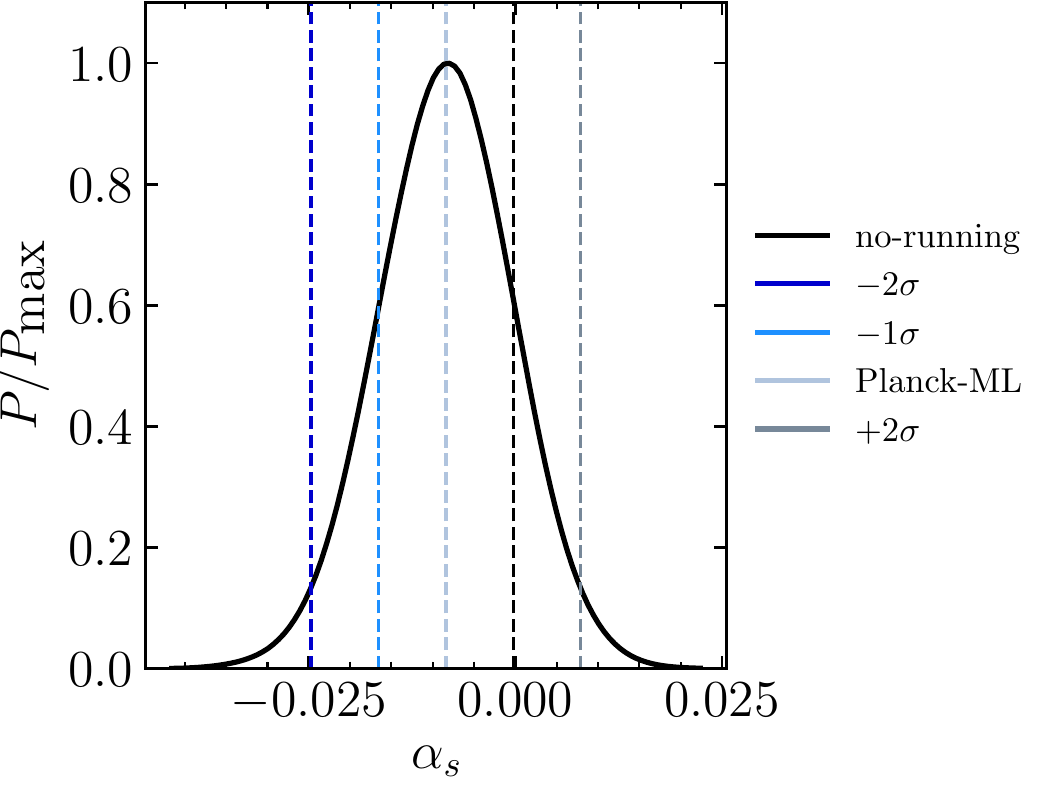}
    \vspace{-0.5cm}
    \caption{Marginalized posterior distribution for the running of the primordial spectral index when included in the $Planck$ MCMC analysis as a free parameter. The vertical coloured lines represent the regions of the $\alpha_s$ distribution function chosen to calculate weighted means for the other important cosmological parameters.}
    \label{fig:1dPD}
\end{figure}

The parameter chains that are used are based on the $Planck$ temperature power spectrum measurements, alongside low multipole measurements of the polarisation power spectrum (\textit{TT+lowTEB}). 
From these, the one-dimensional posterior distribution for $\alpha_s$ was obtained, which can be seen in the top panel of Fig. \ref{fig:1dPD}.  
From this distribution we choose a sample of values for $\alpha_s$ that probe as much of the available parameter space as possible. These values were the maximum likelihood of the distribution, alongside $\pm 1,2 \sigma$ of this value. 
With these adopted values, all chain sets which had a value of the running within $\frac{\sigma}{200}$ of the target value were selected\footnote{This value was chosen so that when the weighted mean of $\alpha_s$ was calculated, it recovered the target value, but still had more than 50 chain sets available for a weighted average of the other important parameters.}. From them, the weighted means of the other important cosmological parameters were taken using the weights of each parameter chain provided. By selecting the values of the other parameters in this way, the predicted angular power spectrum of CMB fluctuations should retain a good match to the $Planck$ data. In other words, we only select `running' cosmologies which are consistent with the observed primary CMB. 

\begin{figure}
    \centering
    \includegraphics[width=\columnwidth]{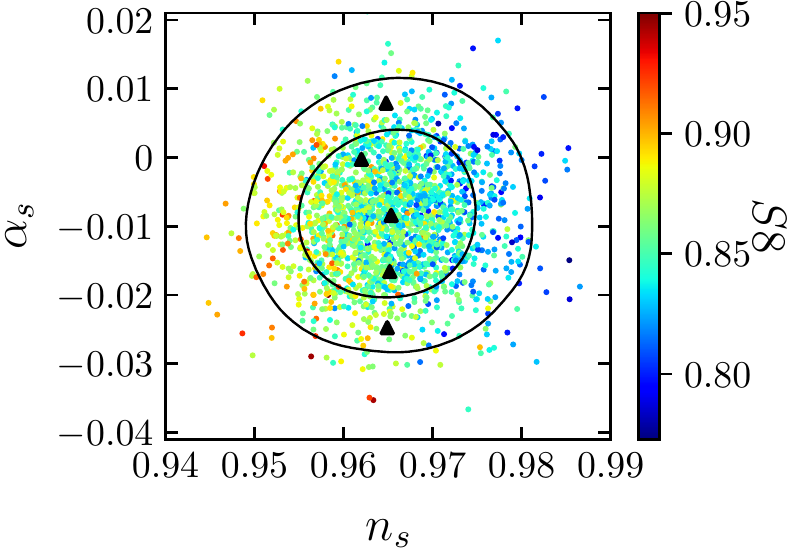}
    \vspace{-0.5cm}
    \caption{Marginalized constraints at 68$\%$ and 95$\%$ CL in the ($n_s-\alpha_s$) plane. Here, points are coloured by that chains' value for the $S_8$ parameter. The black triangles indicate values for ($n_s-\alpha_s$) that were chosen to be simulated.}
    \label{fig:contours}
\end{figure}

That procedure is followed for each chosen value of $\alpha_s$. Note, normally one would also simulate a reference $Planck$ cosmology, with $\alpha_s$ fixed to zero, to be able to quantify the differences between $\Lambda \alpha_s$CDM and $\Lambda$CDM itself. 
However, due to the closeness of the $+1\sigma$ value to $0$ $(\alpha_s = -0.00025)$, this cosmology will be treated as the base $Planck$ no-running cosmology throughout, and will be referred to as the ``no-running" model.
Fig. \ref{fig:contours} shows the 2D marginalized constraints on $n_s$ and $\alpha_s$, with points coloured by the value for the joint constraint on the parameter $S_8 \equiv \sigma_8\sqrt{\frac{\Omega_{\textrm{m}}}{0.3}}$. The resultant cosmologies that were selected can be found in Table \ref{table:cosmoparams} and are indicated by the black triangles in Fig. \ref{fig:contours}. 

\begin{table*}
\caption{The cosmological parameter values for the suite of simulations are presented here. The columns are as follows: (1) The labels for the different cosmologies simulated, that are used throughout the paper. (2) Running of the spectral index, (3) Hubble's constant, (4) present-day baryon density, (5) present-day dark matter density, (6) spectral index, (7) Amplitude of the initial matter power spectrum at a \texttt{CAMB} pivot scale of 0.05 Mpc$^{-1}$, (8) present-day amplitude of the matter power spectrum on scales of 8 Mpc/h in linear theory (note that when computing the initial conditions for 
the simulations, $A_s$ is used, meaning that the ICs are `CMB normalised'). (9) $S_8 \equiv \sigma_8 \sqrt{\Omega_{\rm m}/0.3}$.}
\label{table:cosmoparams}
\begin{tabular}{ccccccccc}
\hline
\multicolumn{1}{l}{(1)} & \multicolumn{1}{l}{(2)} & \multicolumn{1}{l}{(3)} & \multicolumn{1}{l}{(4)} & \multicolumn{1}{l}{(5)} & \multicolumn{1}{l}{(6)} & \multicolumn{1}{l}{(7)} & \multicolumn{1}{l}{(8)} & \multicolumn{1}{l}{(9)} \\ \hline
Label & $\alpha_s$ & $H_0$ & $\Omega_b$ & $\Omega_{\rm CDM}$ & $n_s$ & $A_s$ & $\sigma_8$ & $S_8$ \\
 &  & (km/s/Mpc) &  &  &  & (10$^{-9}$) &  &  \\ \hline
-2$\sigma$ & -0.02473 & 67.53 & 0.04959 & 0.26380 & 0.96201 & 2.34880 & 0.85147 & 0.87286 \\
-1$\sigma$ & -0.01657 & 67.72 & 0.04905 & 0.26018 & 0.96491 & 2.28466 & 0.83939 & 0.85468 \\
$Planck$ ML & -0.00841 & 67.54 & 0.04990 & 0.26186 & 0.96519 & 2.24159 & 0.83442 & 0.85167 \\
no-running & -0.00025 & 67.39 & 0.04901 & 0.26360 & 0.96535 & 2.21052 & 0.83156 & 0.85119 \\
+2$\sigma$ & 0.00791 & 66.95 & 0.04915 & 0.26843 & 0.96478 & 2.14737 & 0.82140 & 0.85013 \\ \hline
\end{tabular}
\end{table*}

As a test, we have verified that when selecting the parameters in this way the resultant predicted CMB TT angular power spectrum (as computed by \texttt{CAMB}) for each of these different cosmologies is consistent with the $Planck$ 2015 angular power spectrum, which they are (see Fig. \ref{fig:ang_power_spec}).
Choosing the parameters in this way, however, has a non-negligible effect on the matter power spectrum (seen in Fig. \ref{fig:Pk}), particularly on the largest and smallest scales, the latter of which are not probed by the $Planck$ data.
By forcing the model to match the CMB angular power spectrum over a range of scales, the amplitude of the matter power spectrum at the pivot scale is forced to vary between the models. 
The result of which is a negative running that has a larger amplitude ($A_s$), and a positive running that has a lower amplitude, compared with the standard no-running model. 
Therefore, it can be expected from the matter power spectrum alone that the inclusion of $\alpha_s$ in the standard model should have measurable effects on the LSS seen in the simulations, with the magnitude and sign of the effect dependent on what range of modes is sampled within the simulated volume.

As briefly discussed in the introduction, a small number of mild `internal' tensions in the $Planck$ CMB analysis have previously been noted (e.g., \citealt{Addison2016,Planck2017LI}) and these could have some bearing on cosmological parameter selection. Of particular relevance for LSS is the apparent asymmetry in the cosmological parameter constraints when derived from low and high multipole ranges \citep{Addison2016}. In particular, the high-$\ell$ peaks and troughs in the observed angular power spectrum appear smoother than that predicted by the best-fit $\Lambda$CDM model. This is qualitatively similar to the effect of lensing of the CMB by LSS, hence when the CMB lensing amplitude, $A_{\textrm{lens}}$, is allowed to float (rather than fixing to the natural value of unity), the CMB TT power spectrum prefers $A_{\textrm{lens}} > 1$. Allowing the lensing amplitude to float results in cosmological parameter constraints that are insensitive to the range of multipoles analysed, but does result in a few sizeable shifts (1-2 sigma) of parameters important for LSS, including $\sigma_8$ and $\Omega_m$ \citep{Addison2016,McCarthy2018}.  

The $Planck$ team did not explore the potential impact of allowing $A_{\textrm{lens}}$ to float on the constraints on the running of the scalar spectral index. In Appendix \ref{appendix:sim_setup} we examine the constraints on the running while marginalizing over $A_{\textrm{lens}}$. We show that the constraints on the running are virtually unaffected by marginalizing over $A_{\textrm{lens}}$.  We therefore adopt the publicly-available $Planck$ 2015 chains with $A_{\textrm{lens}}=1$ for our analyses.

 Lastly, it is worth noting that the effects on the statistics examined in this study should not be solely attributed to a running of the scalar spectral index.  This is because as we vary the running, the other cosmological parameters change in order to retain consistency with the primary CMB constraints from $Planck$. Of particular relevance for this study is the amplitude of the initial matter power spectrum, $A_s$, which we discuss further in Section \ref{sec:matpow}.  As such, when discussing the results from the simulations, we will refer to the changes seen as being a result of a $Planck$-constrained running cosmology, abbreviated as a $\Lambda\alpha_s$CDM cosmology, rather than due to the running parameter alone.

\section{Effects on large-scale structure}

Here we present our predictions for the effects that a  $\Lambda\alpha_s$CDM cosmology has on LSS. Note that the results shown in this section and Section \ref{sec:Internal_effects} are derived from the suite of dark matter only simulations. The effects due to the inclusion of baryonic physics are explored in Section \ref{sec:separability:checks}.

Furthermore, as part of the \texttt{BAHAMAS} project, we are exploring the effects that massive neutrinos \citep[][]{Mummery2017,McCarthy2018} and dynamical dark energy (Pfeifer et al, in prep) have on LSS.  
We will comment throughout on the similarities and differences between these extensions to $\Lambda$CDM.

\label{sec:LSS}

\subsection{Non-linear matter power spectrum}
\label{sec:matpow}
\begin{figure}
    \centering
    \includegraphics[width=\columnwidth]{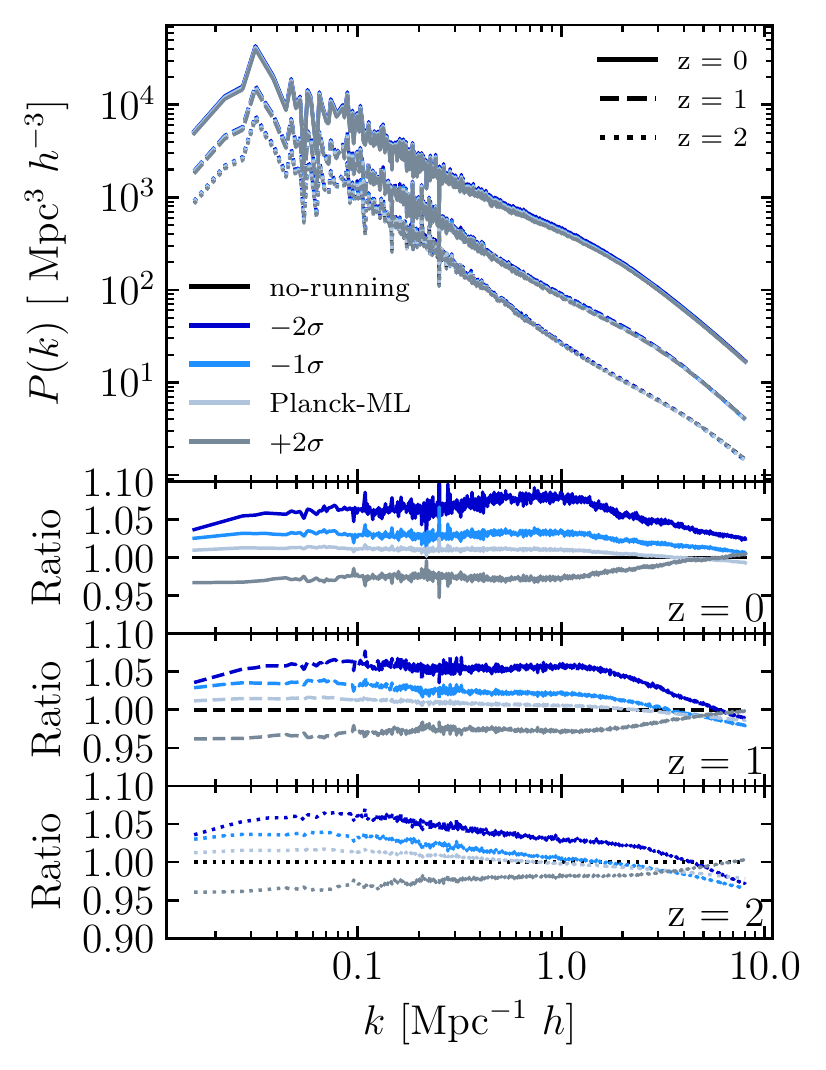}
    \vspace{-0.5cm}
    \caption{Top: the 3D total matter power spectrum for all 5 cosmologies. The different linestyles indicate the matter power spectra measured from the simulations at redshifts 0, 1 and 2 (solid, dashed and dotted respectively). Bottom: the matter power spectra of the different runs normalised to the no-running model result at a given redshift. A negative running cosmology can lead to an amplification of power on large scales by up to 5-10$\%$ compared to a standard $\Lambda$CDM cosmology, and a positive running cosmology can lead to a $\sim$ 5$\%$ reduction in power.}
    \label{fig:Pk_DM}
\end{figure}
 We first look at what effect a cosmology with a free-running spectral index has on the recovered non-linear matter power spectrum extracted from the simulations for redshifts $\leq$ 2.
To compute this, we make use of the publicly-available code \texttt{GenPK}\footnote{\href{https://github.com/sbird/GenPK/}{https://github.com/sbird/GenPK/}} which computes the 3D matter power spectrum for each particle species in the simulation. 
The calculated total matter power spectrum is shown in Fig. \ref{fig:Pk_DM} for $z = 0, 1, 2$. The power spectra are plotted up to a maximum $k$-mode equal to the Nyquist frequency of the simulation: $k_{nyq} \equiv \pi (N/L)$, where $N$ is the cube root of the total number of particles in the simulation, and L, the side length of the box. 
This means that the simulations are able to probe the power spectrum into the linear and non-linear regimes ($k > 0.1$ Mpc$^{-1} h$), allowing us to see the effect a running spectral index has on these scales. 
The bottom panel(s) of this plot shows the matter power spectra of the separate cosmologies, at the three different redshifts, normalised with respect to the no-running case at the corresponding redshift. 
It can be seen that, although there is a large amount of scatter at large scales in the un-normalised power spectra (which arises due to the fact that the simulations do not sample many independent modes on these scales), this scatter largely divides out in the ratios. This is because the ICs of the simulations, as mentioned, have the same random phases (i.e., there is no cosmic variance between the different volumes).

The result shown in the bottom panel(s) is similar to that seen in the linear matter power spectrum (see Fig. \ref{fig:Pk}), in that a  $\Lambda\alpha_s$CDM cosmology which has a negative value for the running produces an excess of power in this $k-$range, and  one which has a positive value leads to a suppression of power.
It is worth noting, however, that these effects (the enhancement of power in the negative running cosmology for example) extend to higher $k$-modes compared with what was seen in the linear matter power spectrum.  The reason for this is that there is a transfer of power from large scales down to small scales during non-linear growth.  As expected, the inclusion of running affects all scales in the simulations, with the maximum effect being seen at $k-$scales around 0.1-1.0 Mpc$^{-1} h$, of $\approx$ 5-10$\%$ increase in power on these scales in the most negative running cosmology, and $\approx$ 5$\%$ suppression in power on these scales in the positive running cosmology. To put this into perspective, both LSST \citep{LSST_white_paper} and Euclid \citep{Amendola2013} are aiming to measure the matter power spectrum (via weak gravitational lensing and galaxy clustering) to a precision of better than 1$\%$ on scales (larger than, and) probed by this volume (see also \citealt{Huterer2002,Huterer2005,Hearin2012}).  If this precision can be achieved, measurements from LSST and Euclid should be able to distinguish between the (CMB-constrained) cosmologies we have explored here.   

Note that there is a slight redshift dependence in the differences between the different cosmologies, with the largest amount of evolution apparent in the $-2\sigma$ cosmology.
This redshift evolution highlights the transfer of power from large scales to small scales.  For example, in the most negative running cosmology, the $k$-scale where there is a transition from an enhancement of power to a suppression, moves to larger $k$-scales (i.e., smaller physical scales) with increasing redshift. 

It is interesting that this cosmological volume size samples the region of the power spectrum which sees a negative running produce an amplification, and a positive running produce a suppression. 
The reason for this being that the $Planck$ pivot scale, and so the scale at which $A_s$ is defined, corresponds to cluster scales, i.e., scales for which the dominant contribution to the matter power spectrum comes from groups and clusters \citep{VanDaalen2015}. These are also the scales \texttt{BAHAMAS} is designed to sample. Thus, because introducing a negative (positive) running into the standard model of cosmology leads to an increase (decrease) in $A_s$ (see Section \ref{ParamSelection}), we see this effect in the power spectrum of the simulations. 
It can be expected from Fig. \ref{fig:Pk} that if the resolution of the simulations were significantly increased or, alternatively, significantly larger volumes were simulated, the effects one might naively associate with a negative, or positive running, i.e. a suppression and amplification of power on these larger and smaller scales respectively, would be more apparent. 

We note that the effects on the power spectrum shown here are of a similar magnitude as those that are seen through the inclusion of (CMB-constrained) dynamical dark energy in the standard model (Pfeifer et al, in prep).  The results seen in the dynamical dark energy extension also display an almost scale independent (over the $k-$range sampled in these studies, at least) suppression/amplification of the power spectrum (depending on the values for $w_0$ and $w_a$), with a maximum magnitude of around 10$\%$.  By constrast, the inclusion of massive neutrinos always leads to a suppression of the matter power spectrum. For example, \cite{Mummery2017} showed that for $\Sigma M_{\nu} = 0.24$ (0.48) eV, the matter power spectrum is suppressed by nearly $15\%$ ($30\%$) on non-linear scales.  (See \citealt{McCarthy2018} for a discussion of current constraints on the summed mass of neutrinos.)

\begin{figure}
    \centering
    \includegraphics{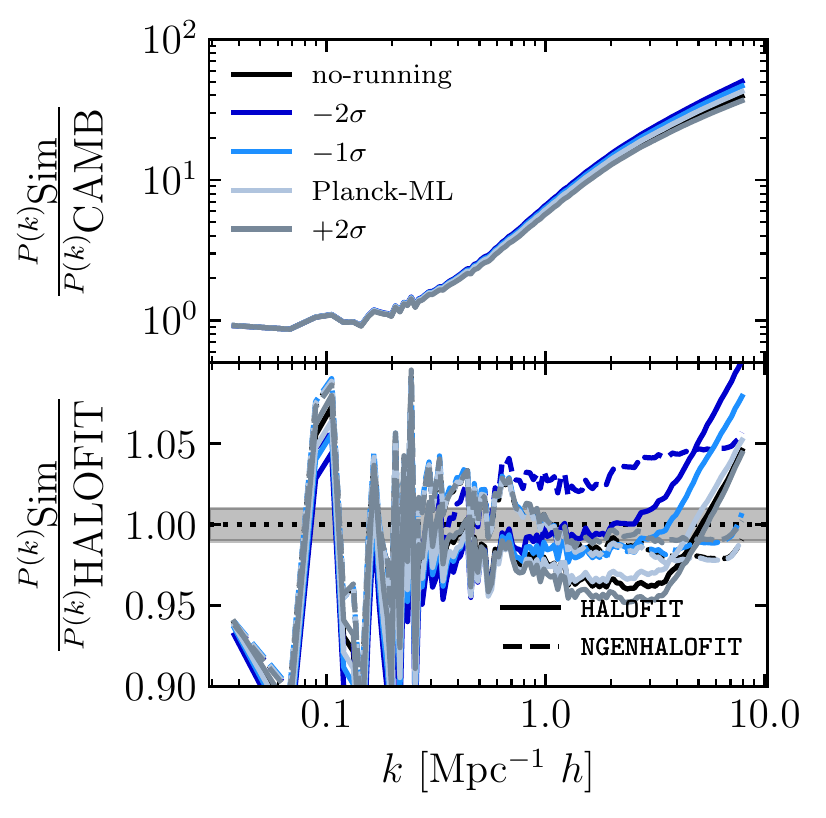}
    \vspace{-0.5cm}
    \caption{Top: the power spectrum output from the simulation at $z = 0$, normalised with respect to the \texttt{CAMB} result at this redshift. This demonstrates the non-linear growth of structure, which linear theory is not able to predict. Bottom: the power spectrum output from the simulation, normalised with respect to the \texttt{HALOFIT} non-linear prediction. A prominent feature present in both panels is the large fluctuations at low $k$ which arise because the simulated volume does not sample these $k-$modes well, leading to increased numerical noise. Also shown is the non-linear power spectrum prediction from \texttt{NGENHALOFIT} (dashed line), which models the non-linear effects of a running scalar spectral index much better, except for the -2$\sigma$ cosmology. This is highlighted by the shaded region centred on unity which represents a 1$\%$ accuracy region. It can be seen that almost all models lie within or just outside this shaded region up to $k$ $\sim$ 5 Mpc$^{-1} h$.}
    \label{fig:Pk_nonlinear_comp}
\end{figure}

In Fig. \ref{fig:Pk_nonlinear_comp} we show the ratio of the non-linear matter power spectrum from the simulations to the linear prediction result from \texttt{CAMB} (top) and the non-linear prediction from \texttt{HALOFIT} \citep{Smith2003, Takahashi2012} (bottom). The top panel isolates the non-linear growth of structure.  We note here that the most negative running case shows the strongest non-linear growth, most likely because this model has an enhancement of power on large scales (due to the increase in $A_s$), which is transferred to small scales during non-linear evolution.  The bottom panel tests the accuracy of the \texttt{HALOFIT} prescription for the non-linear matter power spectrum in running cosmologies. It can be seen that up to $k \approx 4$ Mpc$^{-1} h$, it can reproduce the power spectrum relatively well (to within 5$\%$). However, on scales smaller than this, it appears that \texttt{HALOFIT} does not accurately model the impact that a running scalar spectral index has on the non-linear matter power spectrum. 

However, a recent study into modelling the non-linear effects of a running scalar spectral index on the matter power spectrum was performed by \cite{Smith2019}. This was done using a suite of high-resolution N-body simulations, with an extended cosmological parameter space, with values for these parameters centered on the best-fit $Planck$ 2015 standard model.  \cite{Smith2019} produced a modification to \texttt{HALOFIT} to try to better model the non-linear effects of non-standard cosmologies.   We have used the publicly-available software developed by \cite{Smith2019} \texttt{NGENHALOFIT}\footnote{\href{https://bitbucket.org/ngenhalofitteam/}{https://bitbucket.org/ngenhalofitteam/}} to generate a non-linear matter power spectrum at $z=0$ for the 5 separate cosmologies explored here.  These are also plotted in Fig. \ref{fig:Pk_nonlinear_comp}, shown in the bottom plot as dashed lines. It can be seen that for 3 out of 4 of the different running models \texttt{NGENHALOFIT} does indeed do better at reproducing the non-linear matter power spectrum. There is still an offset for the $-2\sigma$ cosmology, which may be due to this cosmology being more extreme compared to the running cosmologies sampled in \cite{Smith2019} ($\alpha_s = -0.01, 0.01$). 
 Note, as mentioned in Section \ref{sec:Sims}, all of the \texttt{BAHAMAS} simulations include a prescription for massive neutrinos. However, the power spectrum computed by \texttt{GenPK} in the collisionless simulations corresponds to the dark matter + baryons power spectrum (i.e. it does not have a massive neutrino component). As such, when taking the ratio, the corresponding statistic is used in the approximate result (i.e. the power spectra computed by \texttt{HALOFIT}/\texttt{NGENHALOFIT}).

\subsection{Halo counts}
\subsubsection{Halo mass function}\label{section-HMF}
We now examine the effects of a $\Lambda\alpha_s$CDM cosmology on the HMF.  The HMF is defined here as the number of haloes of mass, $M_{200,\textrm{crit}}$, that exist per cubic comoving Mpc, per logarithmic mass interval: $\phi \equiv dn/d\log_{10}(M_{200,\textrm{crit}})$.
The masses of haloes used in this study, unless otherwise stated, represent the mass that is contained within a spherical overdensity whose radius encloses a mean density of 200 times that of the critical density of the Universe at that redshift.
Note also that all distances used in this study are comoving, unless otherwise stated.

Haloes in this study are identified using the \texttt{SUBFIND} algorithm \citep{Springel, Dolag2009} which first runs a standard friends-of-friends (FoF) algorithm on the dark matter distribution, linking all particles which have a separation less than 0.2 $\times$ the mean interparticle separation.  This is used to return the spherical overdensity mass $M_{200,\textrm{crit}}$, it then goes through FoF groups and identifies locally bound sub-structure within each group.  The FoF group is centered on the position of the particle in the central subhalo that has the minimum gravitational potential. 

\begin{figure}
    \centering
    \includegraphics[width=\columnwidth]{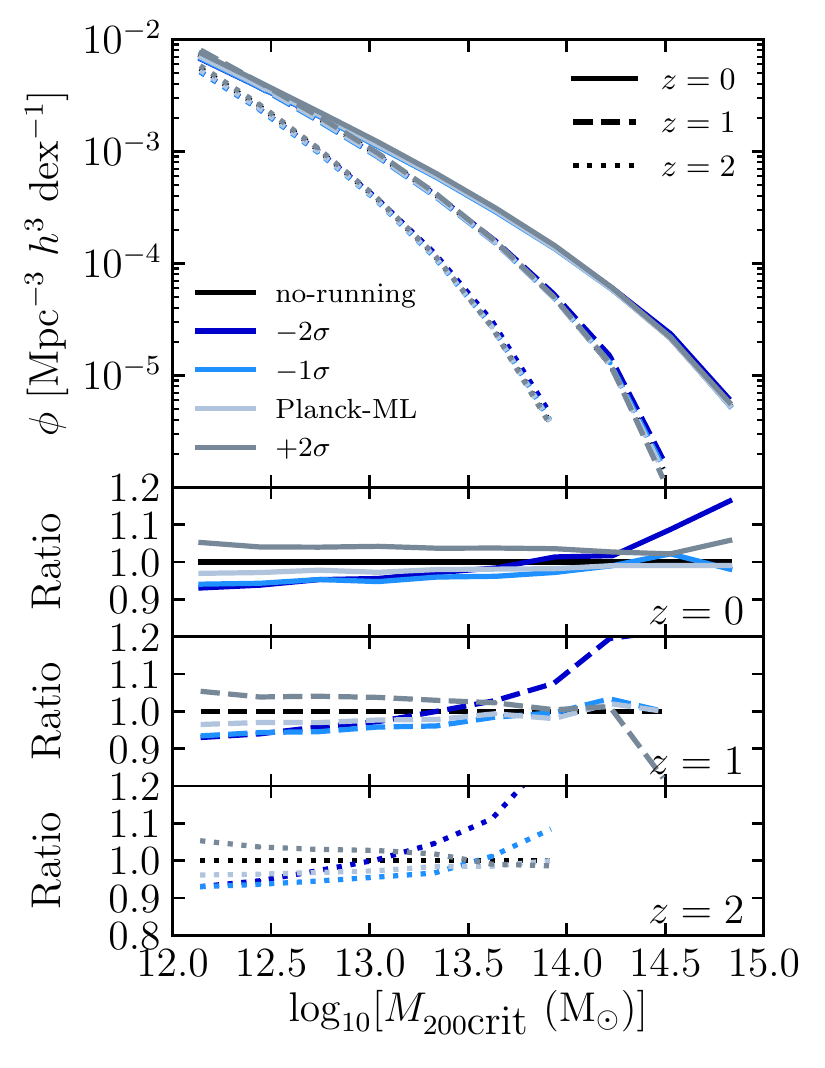}
    \vspace{-0.5cm}
    \caption{Top: Halo mass function (HMF) measured for the 5 separate running cosmologies. The colours here represent the different values for $\alpha_s$, with the linestyles representing the measured HMF at a particular redshift. Bottom: The HMFs normalised with respect to the HMF measured in the no-running simulation. Here the effect due to a change in cosmology and the inclusion of running is evident. A cosmology with a negative running in general leads to fewer low-mass haloes, whereas a cosmology with a positive running leads to more low-mass haloes, with the magnitude of this effect being insensitive to redshift. However, a negative running cosmology predicts more high-mass haloes (at least in the case of the most negative running case), with this effect amplified at earlier redshifts.}
    \label{fig:HMF_DM}
\end{figure}

The measured HMF for the various running simulations can be seen in Fig. \ref{fig:HMF_DM}. The inclusion of running in the simulation has a measureable effect, which is most obvious when looking at the bottom panel of Fig. \ref{fig:HMF_DM}. 
Here the HMF is normalised with respect to the measured HMF in the no-running simulation, and it can be seen that there is an almost 10$\%$ decrease (increase) for the most negative (positive) running cosmology in the number of lower mass haloes that exist in the simulation (10$^{12}$ - 10$^{13}$ M$_{\odot}$).
A similar effect on the HMF in an N-body simulation was found by \cite{Garrison-Kimmel2014}, who showed that the inclusion of a negative running reduced the number of haloes in their simulation at fixed low-mass.  The effect on the HMF due to running depends strongly on the adopted value for $\alpha_s$ as expected, with a more negative value leading to the largest effect on the HMF.
An interesting feature is the fact that the most negative running cosmology appears to predict more haloes that are of a higher mass (10$^{14}$ - 10$^{15}$ M$_{\odot}$), with this effect being stronger at earlier redshifts.
This is likely because these mass scales correspond to the regions of the matter power spectrum where a negative running cosmology leads to an excess of power compared with the no-running model (see Fig. \ref{fig:Pk}). 
Therefore, if the initial seed fluctuations which will grow into these massive haloes are amplified, these haloes will form earlier and will therefore be present at earlier times, compared with the no-running model. 

\begin{figure}
    \centering
    \includegraphics[width=\columnwidth]{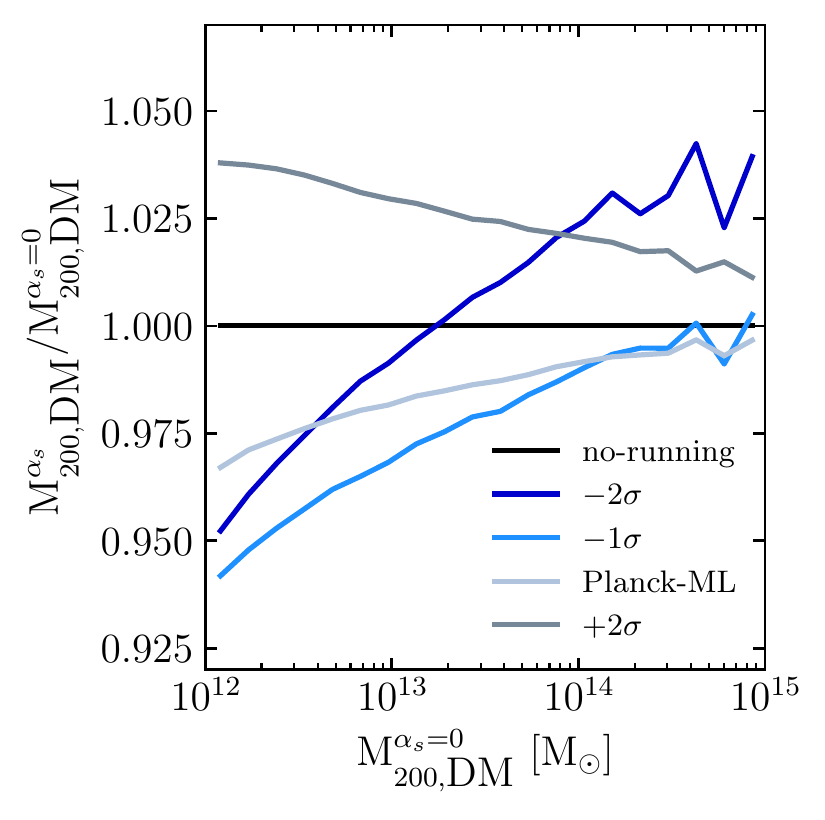}
    \vspace{-0.5cm}
    \caption{The fractional change in halo mass for a matched set of haloes across the 5 different cosmologies, indicated by colour. This plot illustrates the effect a running scalar spectral index has on a halo by halo basis. Here, a halo in a standard $\Lambda$CDM cosmology, would be more massive, if it were instead in a positive running cosmology. Likewise, if it were instead in a negative running cosmology, it could be either less massive or more massive, depending on the magnitude of the running, and the size of the halo. It is this effect on halo mass which drives the differences seen in the HMF.}
    \label{fig:frac_mass_change}
\end{figure}

To try to understand this a little better, it is more intuitive to look at the mass of a halo at fixed number density. 
The reason for this being that it is the halo masses that change, not their number density (i.e. it is a shift along the x-axis, not the y-axis), which leads to the changes seen in the HMF. For example, a halo which evolves from a peak in the density field to a 10$^{15}$M$_{\odot}$ halo will have this density peak either diminished, or enhanced, depending on the value for $\alpha_s$, and thus its final mass is sensitive to this effect.

To look at this, a matched set of haloes needs to be constructed. To match haloes one first needs a reference, and so all haloes in the no-running ($\alpha_s = 0$) dark matter only simulation are chosen as the reference haloes.
For each halo in the reference simulation, a matched halo is found in the simulations with a non-zero value for $\alpha_s$. 
Haloes are matched using the unique particle IDs of the dark matter particles assigned to them. Thus, for each dark matter particle assigned to a halo in the reference simulation, the particle with the matching ID in the other simulations is identified, along with what halo they belong to. 
The halo in each case which contained the largest fraction of identified particles\footnote{Note: a minimum matching threshold is also set, so that a halo must contain at least 51$\%$ of the particles from the reference simulation's halo for it to be classified as the matching halo.} is selected as the matching halo in that simulation. 
Any halo for which a match could not be found across all 5 of the simulations was discarded from the analysis. Overall $>$ 99$\%$ of all haloes in the mass range $10^{12} - 10^{15}$M$_{\odot}$ in the dark matter-only reference simulation were matched across all other simulations. 

The resultant fractional change in halo mass for the matched set of haloes as a function of halo mass in the dark matter only-reference simulation is shown in Fig. \ref{fig:frac_mass_change}.
The result is almost identical to the effect that is seen in the HMF. A cosmology with a positive running leads to more massive haloes across the entire mass range compared with the reference simulation. A cosmology with a negative running generally leads to a decrease in matched halo mass.
This is however a mass-dependent effect, with the more massive haloes not being affected as much in the cosmologies with a negative running.  In fact, for the most negative running cosmology the most massive haloes are somewhat more massive than their no-running counterparts. 
This is what leads to the excess of these haloes in the HMF compared with the standard cosmology. This plot is for haloes at redshift 0, however, we also looked into the fractional change in halo mass at redshifts 1 and 2. It was found that the effects present in HMF are echoed here, in that at earlier redshifts, the effect on halo mass in the two most extreme cosmologies in this study is amplified (Fig. \ref{fig:frac_mass_change_z_evolution}).

Going forward, when using a matched set of haloes, we use the values of $M_{200,\textrm{c}}$ and $R_{200,\textrm{c}}$ for the matching halo in the reference no-running dark matter only simulation. 
\bigskip

When comparing the effects on the HMF due to running, and those due to dynamical dark energy (Pfeifer et al, in prep), and massive neutrinos \citep[][]{Mummery2017}, it is found that massive neutrinos have the largest effect.  For example, for $\Sigma M_{\nu} = 0.24$ (0.48) eV, the number of massive haloes at fixed mass can be suppressed by nearly $20\%$ ($40\%$) at the present day.  This is in comparison to allowing the running to be free but constrained by the CMB, which predicts either no suppression of the most massive objects, or conversely an increase in their numbers.  Dynamical dark energy behaves similarly to a running scalar spectral index, leading to an increase in the number of massive objects at fixed mass (although certain models can lead to a suppression by up to $20 \%$). The effects from massive neutrinos also have the largest redshift dependence, with the effects due to running not varying much with redshift, and those due to dynamical dark energy having a slight redshift dependence, but not to the same extent.

\subsubsection{Comoving halo number density }
\label{sec:comoving_halo_space}
\begin{figure}
    \centering
    \includegraphics[width=\columnwidth]{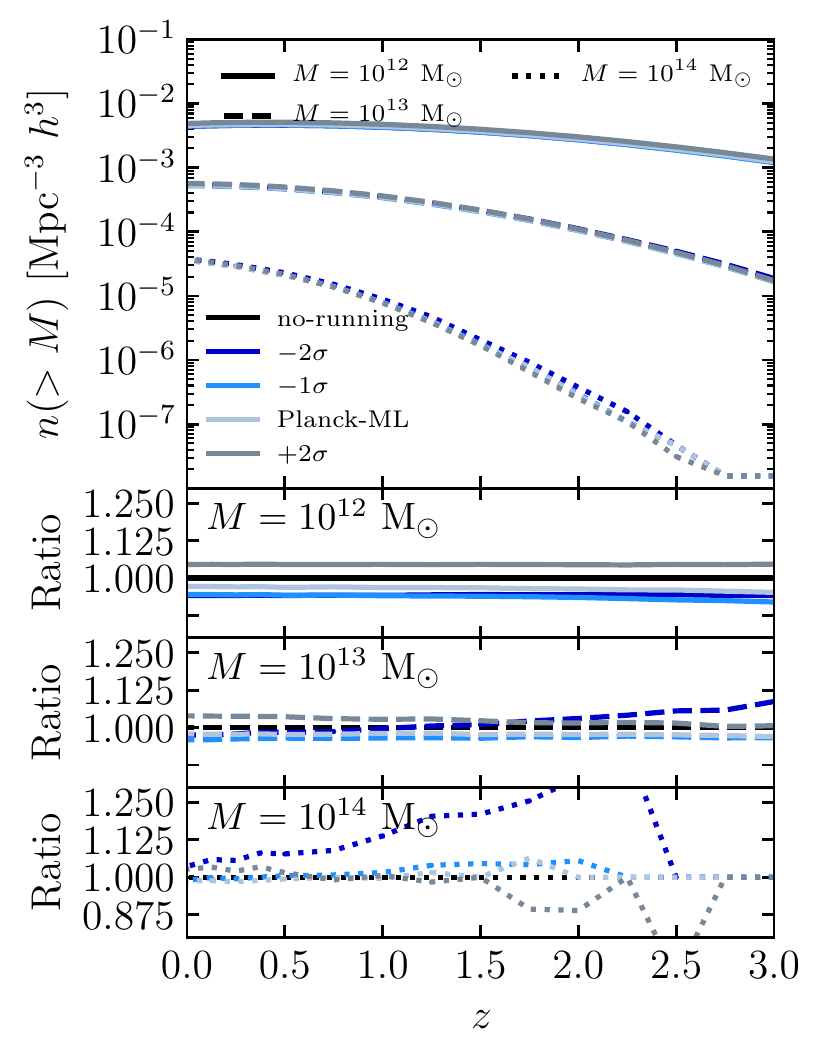}
    \vspace{-0.5cm}
    \caption{Top: redshift evolution of the comoving halo number density, for haloes above three separate mass thresholds. The different linestyles in this plot correspond to the different mass thresholds investigated in this study. Bottom: the comoving halo number density normalised with respect to the no-running model. The effect that a $\Lambda\alpha_s$CDM cosmology has on this quantity is mass and redshift dependent, with for example, the most negative running cosmology having fewer haloes above $10^{12} M_{\odot}$ at all redshifts, but more haloes above $10^{14} M_{\odot}$. This result is amplified with increasing redshift. The opposite holds for the positive running cosmology: more haloes above $10^{12} M_{\odot}$ but fewer below $10^{14} M_{\odot}$.}
    \label{fig:Comoving_Halo_Space_Density}
\end{figure}

The HMF provides a measure of the number density of objects of a certain mass at a certain redshift. Another, similar quantity is the comoving halo number density $n(M,z)$ which is the integral of the HMF above a certain mass threshold, at a certain redshift.
This is a useful quantity to look at, as it is closer to what is actually measured through observations.
The effects that a $\Lambda\alpha_s$CDM cosmology on this quantity are shown in Fig. \ref{fig:Comoving_Halo_Space_Density} for three separate mass thresholds of $10^{12} M_{\odot}$, $10^{13} M_{\odot}$ and $10^{14} M_{\odot}$. Due to the steep, negative slope of the HMF, the majority of counts which make up $n(>M)$ come from haloes closest to the mass cut. 
The bottom panel of Fig. \ref{fig:Comoving_Halo_Space_Density} shows the measured halo number density normalised to the no-running result. It can be seen here that the effects of running are mass dependent, with a negative running  cosmology predicting more massive objects at later redshifts compared with the reference cosmology, and a positive running  cosmology predicting fewer of these massive haloes at later redshifts compared to the negative running and the no-running cosmology.
This result can be understood again in terms of the effect running has on the overdensities in the initial conditions. 
As seen, although a negative running suppresses overdensities on small and large scales, there is a region of the power spectrum which is enhanced  in a best-fit cosmology, with a negative running of the spectral index (due to the simultaneous increase in $A_s$). Thus, these enhanced density perturbations are larger at earlier times than in the reference cosmology, or in a positive running cosmology (which has density perturbations smoothed out on these scales and sees a reduction in power).
As a result, more of these high-mass systems form at earlier times, and because one is looking at rarer systems as the mass threshold is increased, the overall number of objects that exist at early times is small so the relative increase can be quite large. 
\FloatBarrier

\subsection{Clustering of haloes}
\begin{figure}
    \centering
    \includegraphics[width=\columnwidth]{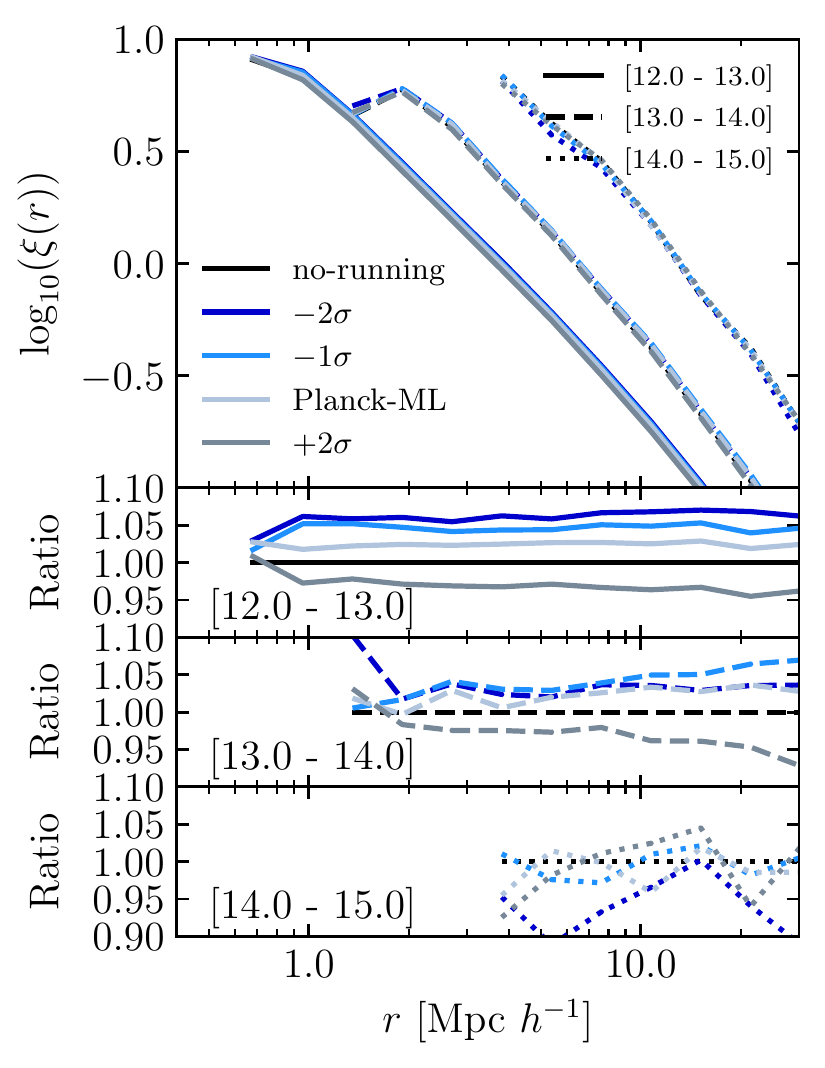}
    \vspace{-0.5cm}
    \caption{Top: the effect a $\Lambda\alpha_s$CDM cosmology has on the 3D two-point halo autocorrelation function, $\xi(r)$. The clustering signal is measured in 3 separate halo mass bins which are indicated by the different line styles. The mass range is shown in the bottom panel and is quoted in units of $\log_{10}(M/\textrm{M}_{\odot})$. Bottom: the correlation function for the different running cosmologies normalised with respect to the no-running model. The introduction of running has a measureable effect on the clustering signal of haloes in the simulation, with the effect depending strongly on mass. For example, the -2$\sigma$ cosmology results in a near 10$\%$ increase in the clustering signal of haloes in the mass range $10^{12} - 10^{13}$M$_{\odot}$, but a near 5$\%$ decrease for haloes in the mass range $10^{14} - 10^{15}$M$_{\odot}$.}
    \label{fig:2PCF}
\end{figure}

Having looked at the effects that including $\alpha_s$ in the cosmological model has on the mass of dark matter haloes, it is interesting to see how it affects the haloes' spatial distribution. In particular, we look at how running affects the 3D two-point autocorrelation function (\citealt{Davis1983}) of haloes. 

The correlation function of matter is related to the power spectrum through the Fourier transform, and dark matter haloes are related to the clustering of matter via a prescription for the halo bias \citep[see][for a recent review]{Desjacques2018}.
Thus, it can be expected based on the analysis of the matter power spectrum that a  $\Lambda\alpha_s$CDM cosmology should have some effect on the spatial distribution of haloes.
In this study we compute the autocorrelation function $\xi(r)$ of FoF groups as the excess probability, compared with a random distribution, of finding another FoF group at some particular comoving distance $r$:
\begin{ceqn}
\begin{align}
    \xi(r) = \frac{DD(r)}{RR(r)}-1,
\end{align}
\end{ceqn}
\noindent where $DD(r)$ and $RR(r)$ are the number of haloes found at a radial distance $r$ in the simulation and the number of haloes expected for a random distribution, respectively.
$RR(r)$ is computed analytically, by assuming that haloes are distributed homogeneously throughout the simulation volume  with a density equal to the mean number density of haloes. 
Furthermore, we compute the correlation function for haloes in specific mass bins, so the mean density of haloes is taken to be the mean density of haloes in a particular mass range. 
To compute $\xi(r)$, we use 20 logarithmically spaced radial bins between 0.1 and 100 comoving Mpc $h^{-1}$.

The effect  that a $\Lambda\alpha_s$CDM cosmology has on the calculated two-point autocorrelation function of haloes in the three separate mass bins is shown in Fig. \ref{fig:2PCF}. It can be seen that the effect depends on halo mass, with the effects most clearly represented in the bottom panel(s) where the measured correlation functions are normalised to the no-running cosmology.
Looking at the most negative running cosmology as an example (as this shows the largest effects), for the lowest mass bin of $10^{12} - 10^{13}$ M$_{\odot}$  there is an overall increase in the amplitude of the correlation function, with the most negative running cosmology predicting an increase of $\approx$ 10$\%$. Whereas for the largest mass bin ($10^{14} - 10^{15}$ M$_{\odot}$) the most negative running cosmology leads to a very mild decrease in the clustering amplitude, with this being around a 5$\%$ decrease.
This makes sense when looking at the HMF, or the comoving halo space density, which showed that a negative running cosmology led to an increase in the number of haloes in this higher mass bin, and therefore one can expect them to be a less biased tracer of the underlying matter distribution, which as a result will lead to a lower clustering signal compared to the no-running cosmology's result. 
This result agrees with that predicted by \cite{Fedeli2010}, who showed that in a negative running cosmology dark matter haloes in the cluster regime are less biased compared with the standard model (see figure 10 in \citealt{Fedeli2010}). 
The other two negative running cosmologies see a similar trend, but not to the same level. Conversely, the positive running cosmology shows the inverse effect, with the clustering amplitude being lower for the lowest mass bin, but slightly increasing as the mass range is increased (although it is still lower, or at the same level, as the no-running model). This again makes sense as the number of objects in the higher mass ranges tends towards the no-running simulation's result (see Fig. \ref{fig:HMF_DM}).

Another feature which is present in Fig. \ref{fig:2PCF} is the downturn in the clustering signal, which occurs at around 0.7, 2, and 4 Mpc $h^{-1}$ for haloes in the mass range: $10^{12} - 10^{13}$ M$_{\odot}$; $10^{13} - 10^{14}$ M$_{\odot}$; $10^{14} - 10^{15}$ M$_{\odot}$ respectively. This downturn is present as these scales correspond to the radius of the FoF haloes in the respective mass bins. On scales smaller than this, FoF haloes overlap and would not be distinguished as separate haloes and so one cannot measure a clustering signal.

So far we examined the clustering signal measured with haloes being placed in mass bins depending on their self-consistent masses, i.e. the mass they have in their own simulation. However, since running changes the mass of a halo (Fig. \ref{fig:frac_mass_change}), it is interesting to look at the effect it has on the distribution of matched haloes, i.e. for haloes of a constant number density.
The reason being that the clustering signal is bound to be different simply because one is looking at a different set of haloes. 
This is done by putting haloes in mass bins based on their no-running cosmology counterpart.
The result on the clustering signal when binning haloes this way is qualitatively the same as that shown in Fig. \ref{fig:2PCF}. Again, in this case a negative running cosmology leads to an increase in the clustering signal of low mass haloes, and a positive running cosmology leads to a decrease, with this effect being mass dependent. The magnitude of this effect also does not change much, with the effect being slightly less in the highest mass bin for the -2$\sigma$ cosmology, but otherwise almost unchanged for the different running models in each mass bin. For brevity we do not show this here.

\bigskip

As a final comparison with the other extensions to $\Lambda$CDM examined in the \texttt{BAHAMAS} project, \cite{Mummery2017} found that the most massive summed neutrino mass cosmology leads to a suppression of the large-scale clustering signal of around 10$\%$, with all non-zero neutrino masses investigated leading to some level of suppression of the clustering signal. Conversely, and similar to what is found in the present study, dynamical dark energy can lead to both a suppression and amplification of the clustering signal, depending on the model. It is also found when comparing the three separate studies that the clustering signal of the largest haloes in the simulations are much less sensitive to the cosmology than the lower-mass haloes. 

\FloatBarrier
\section{Internal Structure of Haloes}
\label{sec:Internal_effects}
Having investigated the effect a running spectral index has on the LSS in the universe, including the abundance of haloes and how they are distributed, we now turn our focus to the internal structure of the haloes themselves. 
To look into this we use two statistics in particular: the spherically-averaged density profiles of haloes, alongside the halo concentration-mass relation. 
\subsection{Total mass density profiles}
\begin{figure*}
    \centering
    \includegraphics[width=\textwidth]{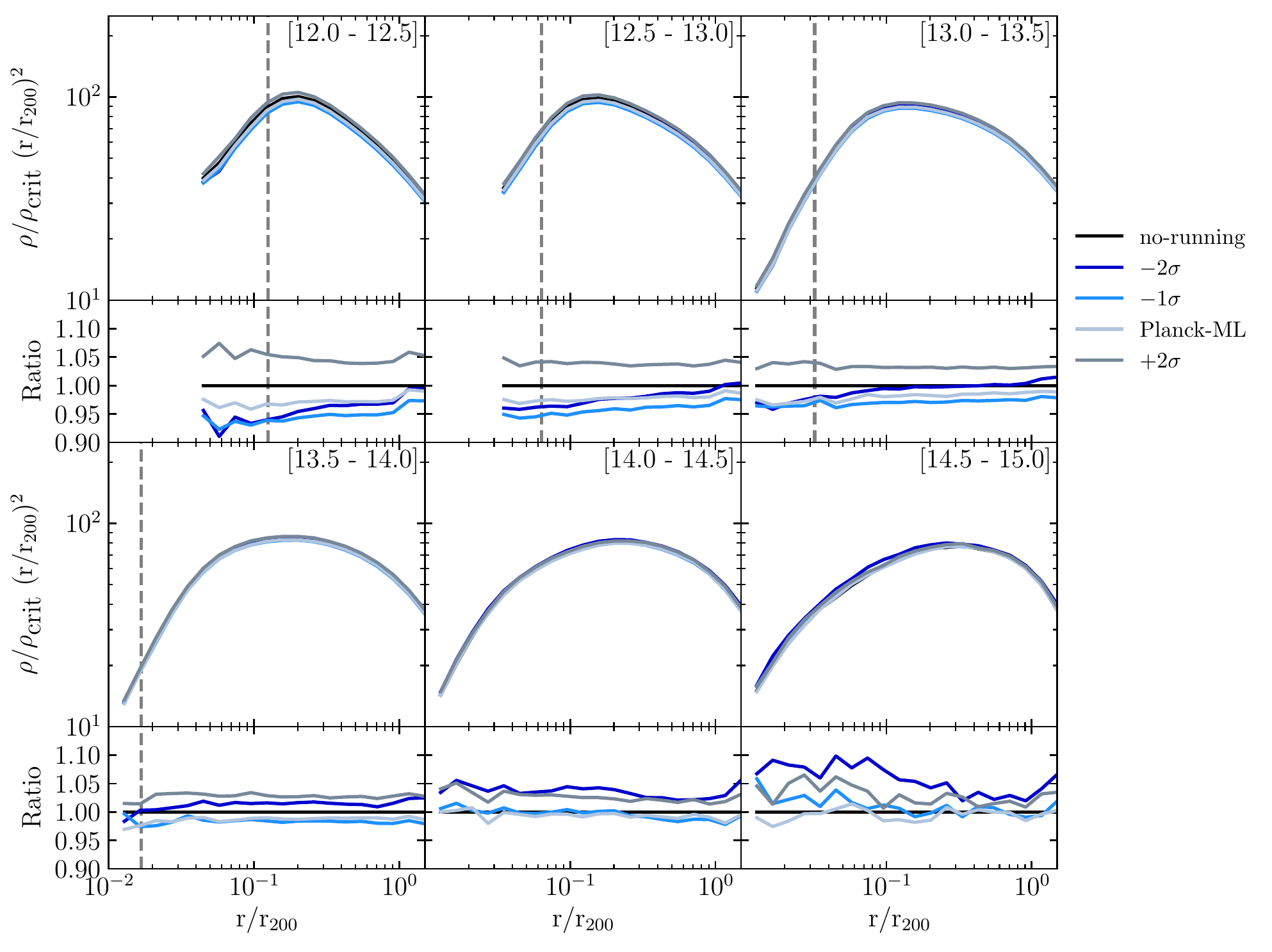}
    \vspace{-0.5cm}
    \caption{Median total mass density profiles of all haloes in each simulation, split into 6 separate mass bins, as indicated in the top right corner of each panel. Haloes are binned according to their mass in the reference no-running cosmology. The halo density profiles are plotted in dimensionless units, normalised by the critical density, and re-scaled so that an isothermal distribution would be a horizontal line. Note, that the profiles are plotted as a function of $r/r_{200}$ also, where $r_{200}$ in this case is taken from the reference simulation. The bottom part of each panel corresponds to the density profiles normalised to that measured in the reference simulation. The vertical dashed line shows the convergence radius (discussed in Section \ref{sec:conc}), and shows the point beyond which the density profiles are converged. The overall effect that a $\Lambda\alpha_s$CDM cosmology has on the density profile of a halo is to either raise or lower its amplitude, without much of an effect on the shape of the profile, although this depends both on mass and on the sign of the running.}
    \label{fig:Density_profiles}
\end{figure*}
To start, we look at the effect on the spherically-averaged total mass density profiles, shown in Fig. \ref{fig:Density_profiles}. 
Each panel shows the median density profile of all haloes in that mass bin (indicated in the top right hand corner of the plot), with the mass bins ranging from $12.0 \leq \log_{10}(M_{200,\textrm{c}}/\textrm{M}_{\odot}) \leq 15.0$, each with a width of 0.5 dex. 
In order to reduce the dynamic range of the plot, we scale the mass density by $r^2$, i.e., so that an isothermal distribution would be a horizontal line. We plot the density profiles in dimensionless units. 
The bottom plot for each panel shows the density profiles normalised to the median density profile of that mass bin, as measured in the reference no-running simulation. 

As mentioned when looking at the two-point autocorrelation function in bins of halo mass, if the haloes are binned according to their self-consistent masses, i.e., the mass they have in their own simulation, this will result in looking at a slightly different population of haloes in each bin across the different simulations.
Instead, the panels in Fig. \ref{fig:Density_profiles} show the median density profile of matched haloes in mass bins corresponding to the masses of haloes in the reference, no-running simulation.
Similarly, the values of $R_{200,\textrm{c}}$ used in this statistic are those corresponding to the haloes in the reference simulations.
This is done as we want to isolate the effects a running scalar spectral index has on a given set of haloes.

It can be seen from Fig. \ref{fig:Density_profiles} that the qualitative effect on the density profiles of haloes in a $\Lambda\alpha_s$CDM cosmology is to either raise or lower the overall amplitude of the density profile, depending on the sign of the running and the mass of the object.
For example, haloes in the lowest mass bin have the amplitude of their density profile decreased in a cosmology which has a negative running and increased in one which as a positive running.
Whereas for haloes in the highest mass bin, almost all cosmologies, regardless of the sign of the running, see an increase in the amplitude of the density profiles.
There is also a hint that a running in this mass range leads to a change in the shape of the density profile, with the central regions being more dense in a cosmology that has a running spectral index.
This trend of a change in amplitude makes sense when looking at Fig. \ref{fig:frac_mass_change}, which showed that for a matched set of haloes a negative running cosmology leads to a decrease in mass for lower-mass haloes but an increase in mass (dependent on the magnitude of $\alpha_s$) for larger-mass haloes.
Conversely, a positive running cosmology led to an increase in mass for all mass ranges.
Also plotted in Fig. \ref{fig:Density_profiles}, shown by the grey-dashed line, is the maximum convergence radius of haloes in each respective mass bin (the convergence radius is discussed in Section \ref{sec:conc}).

\subsection{Concentration-mass relation}
\label{sec:conc}
\begin{figure}
    \centering
    \includegraphics[width=\columnwidth]{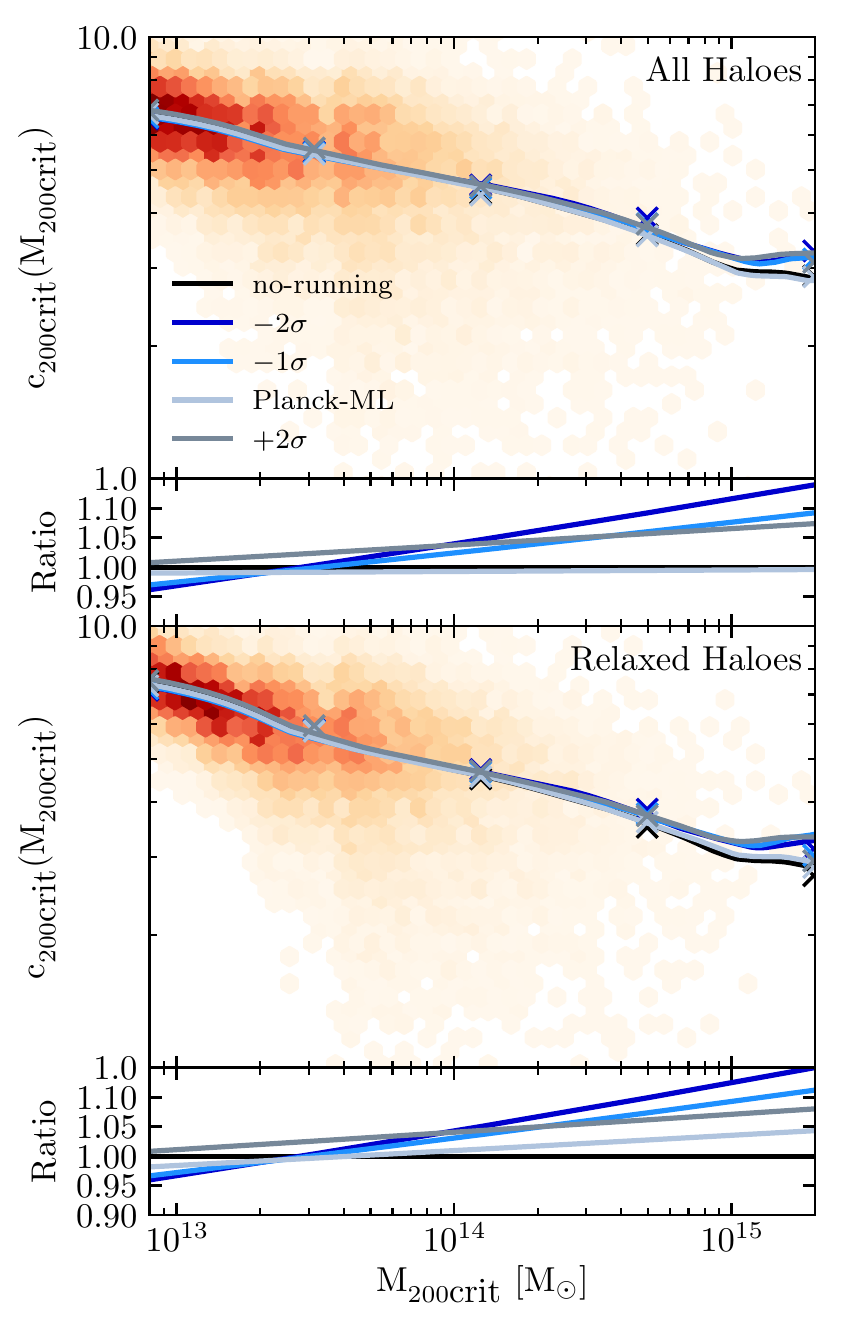}
    \vspace{-0.5cm}  
    \caption{The best-fit c-M relation for all 5 cosmologies simulated. The solid curves represent the running median of the concentrations measured in the 5 separate cosmologies, coloured by their values for $\alpha_s$. 
    The shaded regions represent the scatter that is present in the reference simulation, with the intensity of the colour indicating the number density of haloes in this region. 
    The crosses represent the best fit to the data assuming a power-law relationship. 
    The top half of this plot shows the median c-M relation for all haloes in the simulation above the threshold mass of 10$^{12.8} M_{\odot}$, with the lower subplot showing the Equation \ref{eq:conc} fit normalised with respect to the result obtained in the standard model case. 
    The bottom half of the plot shows the same when only relaxed haloes are included. The inclusion of a running spectral index as a an additional free parameter in the standard model tends to lower the concentration of low-mass haloes, but increase the concentration of high-mass haloes. 
    Making a cut on relaxed haloes reduces the scatter in concentration at fixed mass, however the general result due to a $\Lambda\alpha_s$CDM cosmology is maintained.}
    \label{fig:c-M_relation}
\end{figure}

It has been shown through cosmological simulations that the internal structure of dark matter haloes retain a memory of the conditions of the Universe at the time they were formed, with the formation time of a halo being typically defined as the time when the halo obtains some fraction of its final mass. 
For example, \cite{Navarro1996} showed that lower-mass haloes have a higher central concentration than high-mass ones, which they note is as expected given that lower-mass haloes tend to collapse at a higher redshift when the mean density of the Universe was higher. 
This is a result which has now been confirmed through many N-body simulations, in many different ways \citep[see for example][and references therein]{Child2018}, with many now relating this result to the mass accretion history (MAH) of a halo \citep[e.g.][]{Zhao2003, Correa2015c, Ludlow2016, Child2018}.
The MAH represents the increase in mass of the main progenitor of a halo, with lower mass systems accreting more of their mass at earlier times, while larger haloes assemble most of their mass at later times, when the mean density of the Universe has decreased.

During their hierarchical growth, dark matter haloes have been found to acquire an approximately universal shape described by the Navarro, Frenk $\&$ White density profile (NFW) (\citealt{Navarro1996}), which takes the form of:
\begin{ceqn}
\begin{align}
\label{eq:NFW}
    \rho(r) = \frac{\delta_c \rho_c}{\frac{r}{r_s}\left[1+\frac{r}{r_s}\right]^2},
\end{align}
\end{ceqn}
\noindent where $r$ is the radius, $\delta_c$ is an overdensity parameter, $\rho_c$ is the critical density of the Universe and $r_s$ is the scale radius, corresponding to the radius at which the logarithmic slope of the density profile is -2 (i.e., equal to that of an isothermal distribution).
This profile can equivalently be parameterized with the halo mass $M$ and the halo concentration $c$, which is defined as the ratio of the radius enclosing a spherical overdensity $\Delta$ times the critical density, which in this study we take as 200 times the critical density of the universe, and the scale radius: $c_{200,\textrm{c}} \equiv R_{200,\textrm{c}}/R_s$. 
A result of this is that, if one has a prescription for the concentration-mass (c-M) relation, one can fully specify the internal structure of a DM halo at a fixed mass.

In order to measure the concentration of a halo, we derive an estimate for the scale radius by fitting an NFW profile to each halo in our sample. 
However, haloes are dynamically evolving objects, meaning that when taking these measurements there is the potential that some haloes are not in virial equilibrium and not well described by an NFW profile. 
It has been shown in previous studies that haloes which are not in dynamic equilibrium tend to have lower central densities compared with relaxed haloes \citep[e.g. see][]{Tormen1997, Macci2007, Romano-Diaz}. 
Thus, in order to limit this effect, a simple relaxation test was implemented as proposed by \cite{Neto2007}. 
This deems any halo whose barycentre is offset from the centre of potential by more than 0.07$R_{200,\textrm{c}}$ to be unrelaxed and excluded from our relaxed halo sample. Note that \cite{Neto2007} proposed two further checks on if a halo is relaxed or not, however the test implemented in this current study was shown by \cite{Neto2007} to remove the vast majority of unrelaxed haloes, as such, similar to what was done in \cite{Duffy2008} we only use this criterion to remove unrelaxed haloes.
The centre of mass of a halo is calculated using all of the particles inside $R_{200,\textrm{c}}$ of the halo, and is calculated using the iterative shrinking spheres algorithm (\citealt{Power2003}). 
Another selection criterion we apply to our halo sample is that they must have a minimum of 5000 particles inside $R_{200,\textrm{c}}$. However, due to the relatively low resolution of these simulations, we extend the mass range of our halo samples by stacking haloes that have between $800 \leq N_{200,\textrm{c}} < 5000$ particles, so that the stacked halo has the minimum number of particles required.
This allows us to plot the c-M relation down to haloes of mass $10^{12.8} M_{\odot}$. One final cut on haloes is performed after the NFW profile is fit to the haloes, which sees any halo which has an inferred convergence radius (defined below) $<$ 6 times the gravitational softening length $\epsilon$ removed from the halo sample. 
This is quite a conservative cut and follows \cite{Diemer2015}, who note it was shown that halo density profiles are converged at radii beyond 4-5 $\times$ $\epsilon$ \citep{Klypin2000}.

When fitting an NFW profile to haloes, the radial range used is between $0.1 \leq r/R_{200,\textrm{c}} \leq 1.0$ and is divided into 20 logarithmically spaced bins.
However, as a result this can lead to the innermost bins being dominated by numerical noise, and two-body scattering effects, due to poor particle sampling of these innermost radial bins. This is studied in detail in \cite{Power2003}.
In this study they showed that there is a critical radius below which the density profile of a halo measured in an N-body simulation is not converged (their equation 20, which is satisfied when the collisional relaxation time is roughly the age of the Universe).
Thus, when it comes to fitting an NFW profile to a halo in this study, we only fit to radial bins which lie above the convergence radius. Note that we fit to the quantity: $\rho r^2$ as done in several previous studies (e.g. see \citealt{Neto2007}). Note also that, most recently \cite{Ludlow2019} conducted a convergence study on numerical results from an N-body simulation (such as the shapes of haloes) using haloes simulated inside a cosmological volume, in contrast to the \cite{Power2003} study where they only simulated the convergence of results for a single halo. \cite{Ludlow2019} also provide a convergence criterion, which we tested to see if this resulted in any differences in our computed c-M relation, which it did not. As such, we use the \cite{Power2003} convergence radius when calculating our c-M relation.

The resultant effect a running spectral index has on the c-M relation is shown in Fig. \ref{fig:c-M_relation}.
Here the solid line represents the running median of the concentration, which we calculate using the locally-weighted scatterplot smoothing-method (LOWESS; see \citealt{Cleveland1979}). 
The top panel shows the recovered c-M relation for all haloes in the sample, the bottom panel shows the recovered relation when only using relaxed haloes.

Previous studies have shown that the c-M relation for dark matter only haloes at $z = 0$ is well fitted with a simple power law (e.g. \citealt{Bullock2001}) of the form:
\begin{ceqn}
\begin{align}
\label{eq:conc}
    c_{\Delta} = A\left(\frac{M_{\Delta}}{M_{\textrm{Fiducial}}}\right)^{B},
\end{align}
\end{ceqn}
where we adopt $M_{\textrm{Fiducial}}=10^{14}$M$_{\odot}$.
Equation \ref{eq:conc} is fit to the data, and shown in Fig. \ref{fig:c-M_relation} as crosses. Overall, a power law form is able to describe the c-M relation well.
The bottom panel shown below each main plot shows the result of this fit normalised with respect to the no-running simulation. Also shown by the shaded regions is the overall scatter of the c-M relation at fixed mass, shown just for the no-running cosmology. 
As expected, the scatter is reduced at fixed mass when making a cut on relaxed haloes. 
Looking at the bottom panels it can be seen that there is an overall trend for a $\Lambda\alpha_s$CDM cosmology to produce lower concentrations in low-mass haloes ($M \leq 2 \times 10^{13}$M$_{\odot}$), and larger concentrations in high-mass haloes ($M \geq 2 \times 10^{13}$M$_{\odot}$), although the effect is not as large at low masses as it is at high masses. 

A qualitatively similar result was found by \cite{Fedeli2010}, who showed using semi-analytic methods that the concentration of low-mass haloes in a negative running cosmology is lower than in the standard $\Lambda$CDM model. They also showed that to a small extent this effect was reversed at the high-mass end, with these objects being slightly more concentrated. 
The reason for this may be attributable to formation time. As mentioned, it has been shown that haloes which formed at earlier times have higher concentrations. It has also been shown through the effect on the matter power spectrum that the original overdensities of these high-mass haloes are amplified in cosmologies which have a negative running spectral index. 
As a result, these haloes will have formed earlier on, and thus formed when the universe had a larger mean density, and so compared to the no-running cosmology these objects are centrally denser. 
The fact that the differences seen in concentration at low-masses in the different cosmologies is not as large, compared with that seen at higher masses ($M \approx 10^{15}$M$_{\odot}$), may reflect the fact that because these objects form at even earlier times, their concentrations are less sensitive to their relative formation times. 
\FloatBarrier

\section{Separability of Baryonic Effects}
\label{sec:separability:checks}

We now turn our attention to how separable the effects of allowing the running to be free are from the inclusion of baryonic physics in the simulations.  That is, we explore whether the effects of baryons on LSS are dependent on the choice of running cosmology (and vice-versa).  

The \texttt{BAHAMAS} suite of hydrodynamic simulations are a first attempt at explicitly calibrating the feedback in large-volume cosmological hydrodynamic simulations aiming to quantify the impact of baryon physics on cosmological studies using LSS. 
It is important, however, to check to what extent this calibration is dependent on the cosmology adopted in the simulation. 
The reason being that, if the calibration of the simulations depended significantly upon cosmology, one would have to re-adjust the feedback parameters for each cosmological model simulated. 
Thus, for this reason, the baryonic processes in \texttt{BAHAMAS} were calibrated on internal halo properties (specifically the stellar and baryon fraction of haloes), as opposed to the abundance of haloes or the power spectrum of density fluctuations (for more details see \citealt{McCarthy2017} for the calibration method). 
The benefit of this being that the internal properties of haloes ought to be less sensitive to cosmology. 

With this in mind, we explicitly verified that the stellar and gaseous properties of haloes in the simulations (particularly the galaxy stellar mass function and the gas fractions of groups and clusters) are insensitive to the variations in cosmology presented here. Thus, no aspect of the subgrid physics, feedback or otherwise, was changed from that presented in \cite{McCarthy2017}. 

Previous simulation work has shown that the various physical processes which are involved in galaxy formation and included in modern hydrodynamic simulations as subgrid physics, are capable of affecting the underlying distribution of dark matter. 
For example, it has been shown that the total matter power spectrum (e.g. \citealt{VanDaalen2011}; \citealt{Schneider2015}) and the halo mass function (e.g. \citealt{Sawala2013}; \citealt{Cui2014}; \citealt{Velliscig2014}; \citealt{Cusworth2014}; \citealt{Schaller2015}) along with the binding energies of haloes \citep[][]{Davies2019}, can be affected by a non-negligible amount compared with a dark matter only simulation, through the inclusion of feedback mechanisms.
In addition, this study demonstrates that allowing the running to be free can also have a near 10$\%$ level effect on the total matter power spectrum at $z = 0$ and the halo mass function. 
Hence, an interesting and important question is: how separable are these effects? Can they be treated independently of one another, or do they work to amplify or perhaps suppress certain effects.

To answer this question, we separate the effects into two multiplicative factors: an effect due to the change in cosmology (free-running) and an effect due to baryonic physics. 
This results in the simple ansatz shown below, which is employed when trying to reproduce an observed quantity measured in the full hydrodynamic simulation:
\begin{ceqn}
    \begin{align}
        \label{eq:multi}
        \psi_{\textrm{mult}} = 
        \psi_{\alpha_s = 0}^{\textrm{DM}}
        \left(\frac{\psi_{\alpha_s}^{\textrm{DM}}}{\psi_{\alpha_s = 0}^{\textrm{DM}}}\right)
        \left(\frac{\psi_{\alpha_s = 0}^{\textrm{H}}}{\psi_{\alpha_s = 0}^{\textrm{DM}}}\right),
    \end{align}
\end{ceqn}
\noindent where $\psi$ represents the quantity that is being measured, for example the halo mass function or the matter power spectrum, with $\psi_{\textrm{mult}}$ being the multiplicative prediction made from treating the two effects separately; $\psi^{\textrm{DM}}$ and $\psi^{\textrm{H}}$ represent the quantity measured in the dark matter only, or the full hydrodynamic simulation respectively; $\psi_{\alpha_s = 0}$ and $\psi_{\alpha_s}$ represent the values measured in the simulation which has zero running and a running equal to $\alpha_s$, respectively. 
This equation can be split up into two main parts: one which accounts for the effects due to the change in cosmology (free-running), accounted for by the first bracketed quantity in Equation \ref{eq:multi}; and one which accounts for effects due to baryonic physics, such as AGN feedback, accounted for by the second bracketed term. 

All of the statistics examined in previous sections for the dark matter only simulation were treated with this simple ansatz, to see if the result obtained in the full hydrodynamic simulation could be recovered. However for brevity, we focus our attention on 4 main statistics: the total matter power spectrum, the halo mass function, the two-point halo autocorrelation function, and the density profiles of haloes. 
\FloatBarrier
\subsection{Matter power spectrum}
\begin{figure}
    \centering
    \includegraphics[width=\columnwidth]{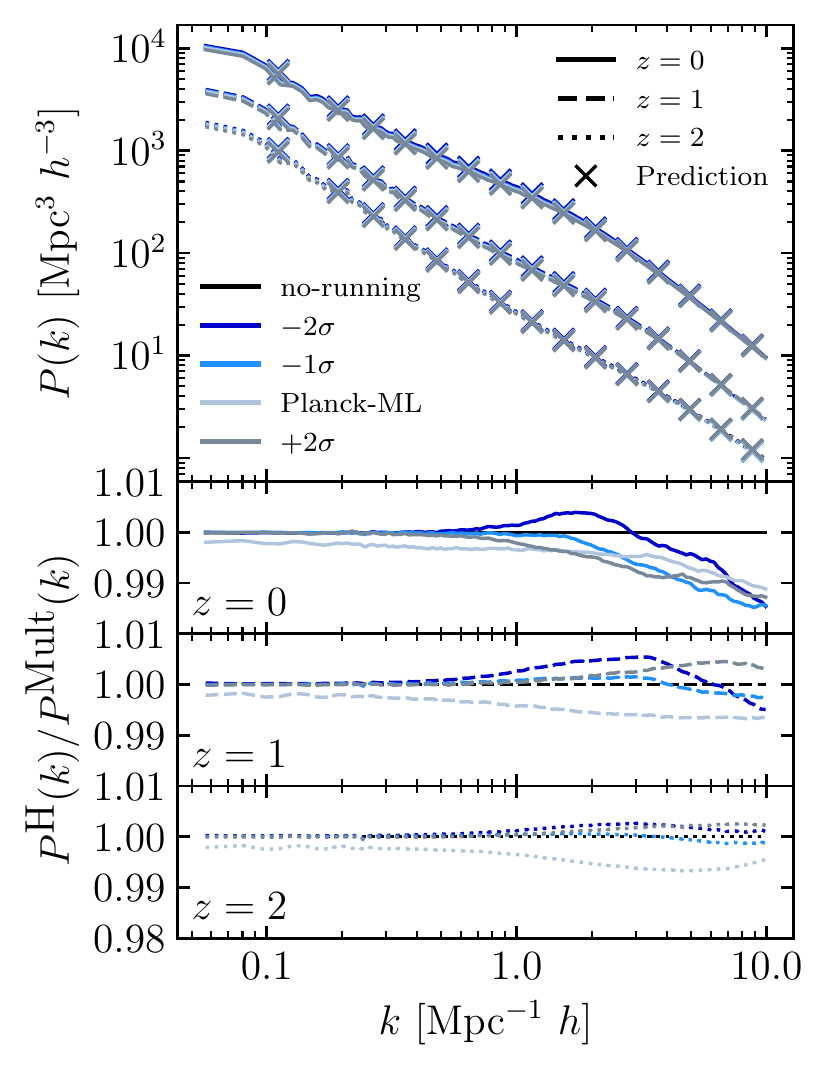}
    \vspace{-0.5cm}
    \caption{Test of the separability of the effects that a running scalar spectral index and baryonic physics have on the total matter power spectrum. Top: The total matter power spectrum for the 5 cosmologies. Here lines represent the result obtained from the simulations, with linestyle representing the different redshifts at which the total matter power spectrum was measured. The crosses represent the result obtained from treating the effects due to a running spectral index and the effects due to baryonic physics on the total matter power spectrum separately. Bottom: the ratio of the self-consistent result measured from the hydro simulation and the result obtained from treating the effects separately. Treating each effect as a simple multiplicative correction, reproduces the total matter power spectrum on all $k-$scales examined here to better than 2$\%$.}
    \label{fig:Pk_HYDRO}
\end{figure}

To begin, we go back to the total matter power spectrum of the simulations.
As described previously (Section \ref{sec:matpow}), the power spectrum is computed using the algorithm \texttt{GenPK} which computes the power spectrum of a simulation snapshot for each individual particle species.
As a result, one first needs to combine the individual matter power components of each particle species to compute a total matter power spectrum.
The resulting total matter power spectrum for the hydrodynamic simulations can be seen in the top panel of Fig. \ref{fig:Pk_HYDRO}.
Here the lines represent the total matter power spectrum measured for 3 different redshifts: $z = 0, 1, 2$. The crosses indicate the recovered result when using Equation \ref{eq:multi} (note, here $\psi = P(k)$) to try to reproduce the result from simultaneously simulating baryonic physics and a running scalar spectral index by treating the effects separately.
The bottom panel(s) of this figure show that over the entire $k-$range examined here, the total matter power spectrum for redshifts out to at least $z = 2$ can be reproduced to $< 2 \%$ by treating these effects separately. On $k-$scales below 3.0 Mpc$^{-1} h$, the result is even more accurate, with the matter power spectrum being reproduced to sub-percent accuracy.  This is better than the accuracy required for the likes of LSST \citep[e.g.][]{Huterer2005, Hearin2012}.

\subsection{Halo mass function}
\begin{figure}
    \centering
    \includegraphics[width=\columnwidth]{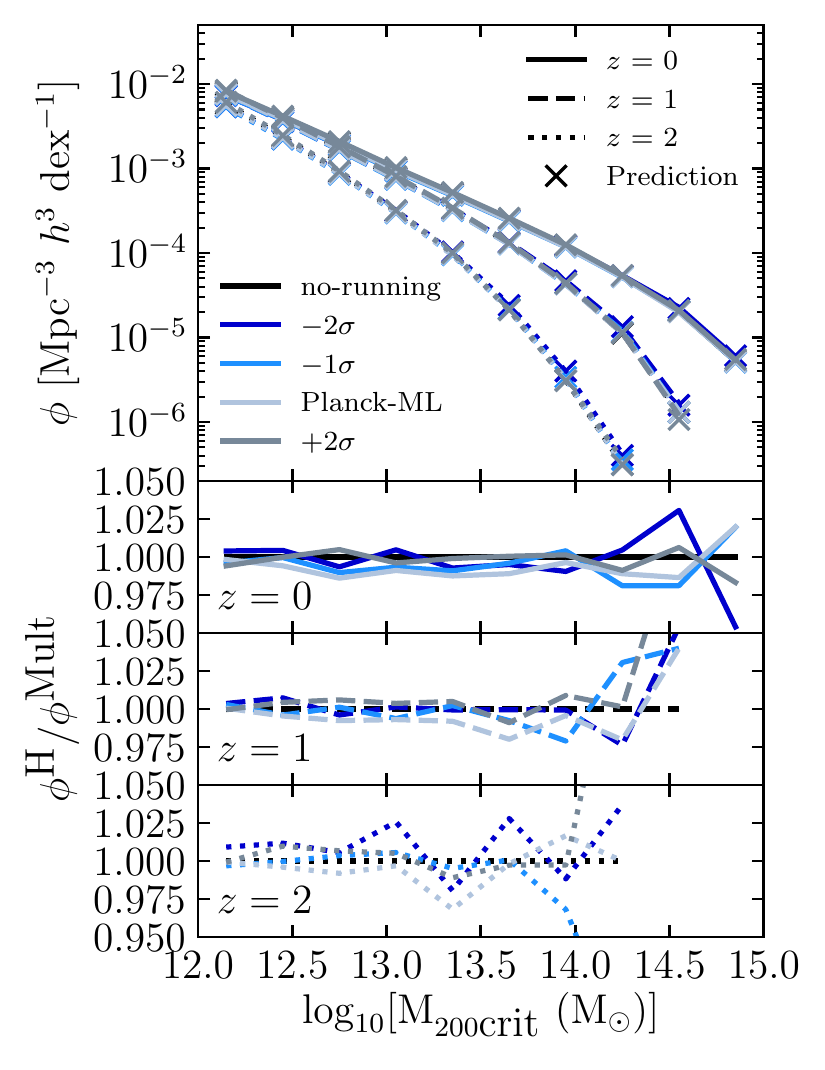}
    \vspace{-0.5cm}
    \caption{Test of the separability of the effects of inclusion of baryonic physics and a running scalar spectral index on the halo mass function measured in a full hydrodynamic simulation. Top: the HMFs measured when simulating both a running spectral index and baryonic physics, with the different linestyles indicating the HMFs measured at different redshifts. The crosses represent the HMF recovered when treating the effects separately. Bottom: the ratio of the measured HMF from the simulation to that predicted using the multiplicative prescription (Equation \ref{eq:multi}). Treating each effect as a multiplicative correction is able to reproduce the measured HMF to better than $\approx 3\%$ up to $z = 2$.}
    \label{fig:HMF_HYDRO}
\end{figure}

Next, we look at the HMF, which is shown in Fig. \ref{fig:HMF_HYDRO}. The HMF as measured for the 5 separate cosmologies in the full hydro simulation is shown by the different lines and is plotted for redshifts: $z = 0, 1, 2$. 
We use Equation \ref{eq:multi} (where $\psi \equiv \phi$ in this case) to test how separable the effects on the HMF due to the inclusion of baryonic physics and a running scalar spectral index are.
The resultant multiplicative prediction is shown as crosses in Fig. \ref{fig:HMF_HYDRO}. The bottom panel(s) of this figure shows the ratio for each cosmology of the measured result from the hydro simulation, i.e. treating both a running scalar spectral index and baryonic physics at the same time, to the predicted result from the multiplicative prescription treating each effect separately. 
It can be seen that for all redshifts examined here, the HMF can be reproduced with this simple ansatz to within better than $\approx$ 3$\%$ over the full range of halo masses examined in this study, up to $z = 2$, with an even better accuracy for lower redshifts.

As a separate test, we look at how separable these effects are on the mass of a halo, with baryonic physics known to reduce the mass of a dark matter halo in the mass range examined in this study \citep[e.g.][]{Sawala2013, Cui2014, Velliscig2014, Schaller2015}, and having seen here that a running spectral index can also have a significant effect on halo mass. 

\begin{figure}
    \centering
    \includegraphics[width=\columnwidth]{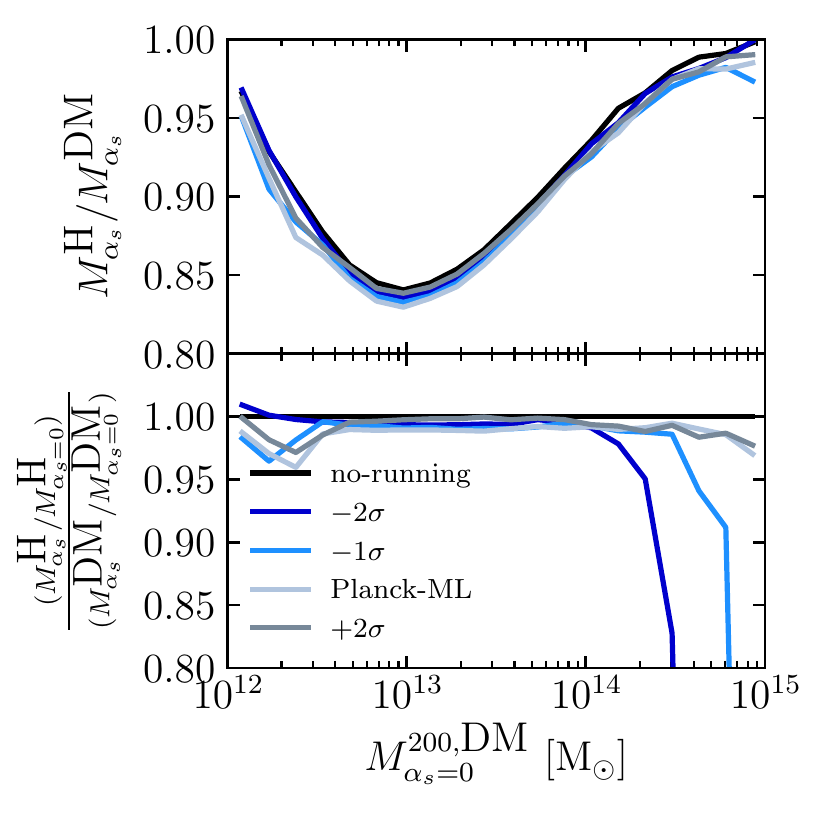}
    \vspace{-0.5cm}
    \caption{Test of the separability of the effect of baryonic physics and a running scalar spectral index on the mass of haloes. Top: the reduction in halo mass of a matched set of haloes due the inclusion of baryonic physics. Bottom: the effect a running scalar spectral index has on halo mass in the full \texttt{BAHAMAS} hydrodynamic simulations, normalised to the collisionless case. This shows that for most of the mass range sampled here the effect on halo mass due to a running spectral index is insensitive to the implemented baryonic physics.}
    \label{fig:frac_mass_change_HYDRO}
\end{figure}

The result is shown in Fig. \ref{fig:frac_mass_change_HYDRO}. The top panel shows the effect baryonic physics has on halo mass, for different values of the running scalar spectral index, i.e. it shows, for a fixed value of $\alpha_s$, the ratio of masses of a matched set of haloes in the hydrodynamic and dark matter only simulations. 
The result shown is the median ratio in bins of halo mass. This shows that, to an accuracy of $\approx$ $1\%$, the effect on halo mass due to baryonic physics is independent of the change in cosmology. 
Also shown in the bottom panel of this figure, is the effect that a running scalar spectral index has on halo mass in the \texttt{BAHAMAS} calibrated feedback model normalised to the corresponding result from the dark matter only simulations. This allows us to test how sensitive the effect a running scalar spectral index has on halo mass is to different physics models.
It can be seen again here, for almost the entire range in mass, the effect due to a running scalar spectral index is nearly independent of the baryonic physics.  Poor statistics are likely responsible for the larger deviations at high masses, whereas the small deviation at the lowest masses is likely a resolution effect (note that the different cosmologies have slightly different particle masses, due to the different values of $\Omega_{\textrm{m}}$ and $H_0$).

\subsection{Clustering of haloes}
\begin{figure}
    \centering
    \includegraphics[width=\columnwidth]{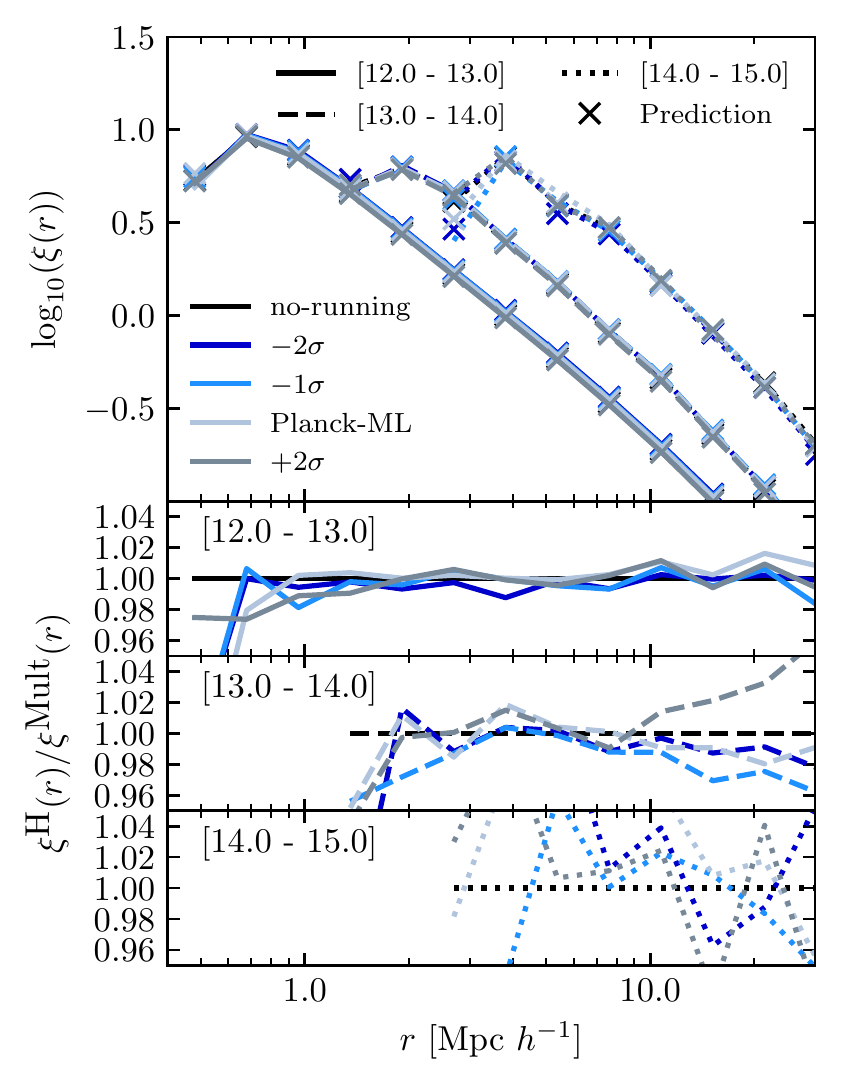}
    \vspace{-0.5cm}
    \caption{Test of the separability of the effects of baryonic physics and a running scalar spectral index on the clustering of haloes in bins of halo mass. Top: clustering signal when simultaneously simulating a running scalar spectral index and baryonic physics. The different linestyles represent the measured clustering between objects in different mass ranges, note that here haloes are binned according to the self-consistent masses of each simulation. The crosses represent the recovered result when treating the effects as separate multiplicative corrections (Equation \ref{eq:multi}). Bottom: the measured clustering signal from simultaneously simulating both effects, normalised with respect to the corresponding multiplicative prediction. For almost all of the radial range examined here, and for all mass bins, the clustering signal is reproduced to better than $\approx 4\%$ by treating these two effects separately.}
    \label{fig:2PCF_HYDRO}
\end{figure}
The separability of the effects due to baryonic physics and allowing the scalar spectral index to be free on the two-point autocorrelation function are shown in Fig. \ref{fig:2PCF_HYDRO}. 
Here haloes are binned using their self-consistent masses, and not the masses of their matched reference counterparts. 
The lines represent the measured two-point autocorrelation function obtained from simulating baryonic physics and a running spectral index simultaneously. The crosses represent the result obtained from using Equation \ref{eq:multi} and multiplying the effects due to each separately (note that in this case: $\psi \equiv \xi(r)$). 
The bottom panel(s) shows the measured result from simulating both effects simultaneously normalised with the result obtained from multiplying the effects separately.
It can be seen that the clustering signal of haloes in separate mass bins can be recovered to better than $\approx 4 \%$ even for the most massive haloes, with the differences being exacerbated due to the low number density of these massive haloes in the sample. 

\subsection{Density profiles}
\begin{figure*}
    \centering
    \includegraphics[width=\textwidth]{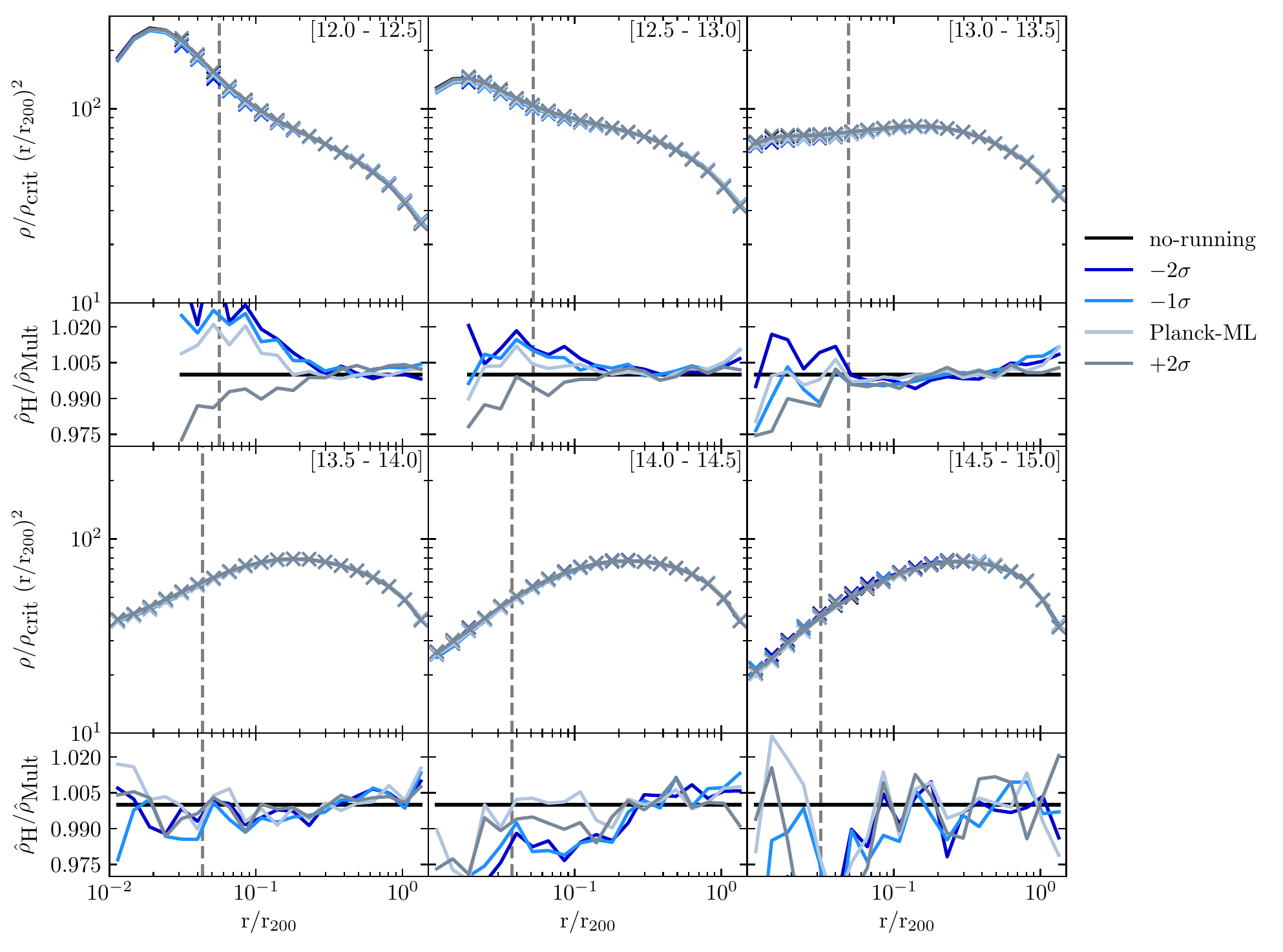}
    \vspace{-0.5cm}
    \caption{Comparison of the median spherically-averaged total-matter halo density profile measured from the hydro simulations, represented by the solid lines in this plot, and calculated using the simple functional form of Equation \ref{eq:multi}, shown by crosses. 
    Haloes are binned using their self-consistent masses, and not that of their matched reference counterparts. The mass-range of each bin is indicated in the top right corner of each panel and is in units of: $\log(M_{200,\textrm{c}}/$M$_{\odot})$. 
    The vertical dashed line in each panel represents the convergence radius for haloes in the reference simulation. 
    The bottom sub-panels show the ratio of the measured self-consistent density profile to that predicted treating both effects separately. The results agree to better than 2$\%$ for the entire radial range for each mass bin.}
    \label{fig:Density_Profiles_HYDRO}
\end{figure*}

Finally, we examine the separability of the effects due to baryonic physics and allowing the running to be free on the total mass density profile of a halo.
As was seen in Fig. \ref{fig:frac_mass_change_HYDRO}, the effects on halo mass due to baryonic physics and the effect on halo mass due to a running spectral index are largely independent effects. As a result, one might expect that the density profiles of these haloes can also be reproduced relatively well with a simple multiplicative prescription.
To test this, we use Equation \ref{eq:multi} (note that here $\psi = \hat{\rho}$ where $\hat{\rho}$ is the dimensionless re-scaled $\rho$: $\hat{\rho} \equiv (\rho/\rho_{\textrm{crit}})(r/R_{200,\textrm{c}})^2$). 
The results are shown in Fig. \ref{fig:Density_Profiles_HYDRO}. 
An important note here is that, unlike in Fig. \ref{fig:Density_profiles}, here we plot the density profiles of an unmatched set of haloes, meaning that the haloes are binned using their self-consistent masses.
Here the lines represent the median spherically averaged total-mass density profiles of each mass bin. The crosses represent the density profiles obtained from treating the effects due to a running scalar spectral index on the density profile of a halo and the effects due to baryonic physics separately. 
The bottom sub-panels show the ratio of the measured density profiles from the simulations to the self-consistent result obtained from treating the effects separately. 
The plot shows that even the combined effects on the internal structure of haloes can be reproduced (to typically better than $2 \%$) by treating baryons and running separately. 

\bigskip

As mentioned, all of the other statistics examined in the dark matter only simulation were also investigated in the way that has just been laid out in this section to see if the effects of baryonic physics and allowing the running to be free could be treated as separate effects.
This includes the comoving halo number density and the c-M relation.
Although these are not shown, they were reproduced using the simple functional form of Equation \ref{eq:multi} to a similar level of accuracy to the tests presented above.

\section{Summary and Conclusions}
\label{sec:conclusions}
This study has made use of a new extension to the \texttt{BAHAMAS} suite of large cosmological hydrodynamic simulations which are 400 comoving Mpc $h^{-1}$ on a side. This extension consists of 5 simulations, corresponding to different values for a running scalar spectral index ($\alpha_s$), ranging from $-0.02473 \leq \alpha_s \leq 0.00791$ (a complete list of values can be found in Table \ref{table:cosmoparams}). 
This has allowed an investigation into the effects a cosmology with a freely varying $\alpha_s$ has on the cosmic LSS and into how separable these effects are from those produced by baryonic physics. 
The statistics which this study focused on include  the non-linear total matter power spectrum, the halo mass function (HMF), the two-point halo autocorrelation function, the spherically averaged halo density profiles, and the halo concentration-mass (c-M) relation. This study focuses mainly on haloes in the mass range: $10^{12} \leq M/\textrm{M}_{\odot} \leq 10^{15}$.
The main conclusions that can be drawn from this paper are as follows:
\begin{itemize}
    \item On almost all scales probed in these simulations, a  cosmology with a negative  value for the running of the scalar spectral index leads to an excess of power compared with the standard cosmology, by a maximum of $\approx10\%$. 
    This holds up to at least $z = 2$ (see Fig. \ref{fig:Pk_DM}).
    Conversely, a  cosmology with a positive running leads to a suppression of the matter power spectrum over the entire $k-$range investigated in this study ($0.01 < k [\textrm{Mpc}^{-1} h] < 8.0$) at $z=0$.  Such an effect should be measurable in upcoming LSS surveys, such as those from LSST and Euclid which aim to measure the power spectrum to around 1\% accuracy. Naively, one might have expected a negative (positive) running to lead to less (more) power relative to a no-running case.  However, the imposed requirement that our adopted cosmologies retain a good fit to the CMB temperature angular power spectrum in the presence of running forces the power spectrum amplitude, $A_s$, to slightly increase (decrease) for cosmologies with negative (positive) running (see Table \ref{table:cosmoparams}).  This effect is retained down to $z=0$ in the non-linear power spectrum.
    \item A cosmology with a running scalar spectral index is capable of suppressing the low-mass end ($M \lesssim 10^{13}$M$_{\odot}$) of the HMF by nearly 10$\%$ in the case of a cosmology with a negative running, but leads to an excess of low-mass haloes for a positive running cosmology (see Fig. \ref{fig:HMF_DM}).
    This effect on the HMF due to a running scalar spectral index depends both on redshift and the mass bin, with a negative running cosmology leading to an excess of high-mass objects ($M \gtrsim 2 \times 10^{13}$M$_{\odot}$) with this effect being pronounced at earlier redshifts.
    This result is due to the fact that, in a negative running cosmology, the density perturbations which lead to these haloes are amplified and thus lead to higher-mass haloes forming earlier on. 
    \item A cosmology with a running spectral index leads to a change in the spatial distribution of haloes (see Fig. \ref{fig:2PCF}). A negative running cosmology leads to haloes with $M < 10^{14}$M$_{\odot}$ being more clustered compared with a standard cosmology (with a near 10$\%$ increase in the clustering signal for the most negative running cosmology) and higher masses being less clustered. 
    This result is reversed for a positive running cosmology, which sees a reduction in the clustering signal of haloes with $M < 10^{14}$M$_{\odot}$. 
    This is due to the fact that, in a negative running cosmology, the formation of these low-mass haloes is suppressed, leading them to be a more biased tracer of the underlying matter distribution, and thus have a stronger clustering signal. 
    \item Looking at internal properties of matched haloes, a cosmology with a running spectral index predominantly causes an amplitude shift in the density profiles of haloes, without much of a change to their shape (see Fig. \ref{fig:Density_profiles}). 
    The effect of a running scalar spectral index depends both on the mass of the halo and the value of the running.
    A cosmology with a negative running leads to a reduction in the amplitude of the density profile for haloes with $M < 10^{13.5}$M$_{\odot}$, but lead to an increase in the amplitude for high-mass ($M \gtrsim 10^{13.5}$M$_{\odot}$) haloes.
    Conversely a cosmology with a positive running leads to an amplitude increase across the entire mass range investigated here.
    This is due to the fact that in a negative running cosmology a low-mass halo has its mass decreased, compared with its matched no-running counterpart. Whereas for the same cosmology, the mass of a high-mass halo (depending on the magnitude of $\alpha_s$) is increased.
    Conversely, a cosmology with a positive $\alpha_s$ leads to an increase in mass of a halo across the entire mass range studied here. 
    \item For the c-M relation, it was found that the effect of a $\Lambda\alpha_s$CDM cosmology is once again mass-dependent, with all running cosmologies predicting the higher-mass haloes to be more centrally concentrated than in a no-running cosmology (see Fig. \ref{fig:c-M_relation}). 
    Whereas at the low-mass end ($M \lesssim 2\times10^{13}$M$_{\odot}$), the negative running cosmologies tend to predict slightly lower concentrations, the effect on these scales is small compared to the high-mass end. 
    A similar result was found by \cite{Fedeli2010}, who predicted low-mass haloes to have a lower concentration value in a negative running cosmology, but high-mass haloes to have a slightly higher concentration. 
    
    Note that in the calculation of the c-M relation, haloes are binned according to their self-consistent masses, not the masses of their no-running counterparts (as was the case when looking at the density profiles). This is the reason why we see some differences in the c-M relation, with these differences not being obvious from looking at the density profiles of the matched haloes. 
 
    \item This study also looked into the separability of the effects due to a running scalar spectral index and baryonic physics.
    We found that these two effects can be treated as separable multiplicative corrections to a standard cosmology.
    This simple multiplicative procedure is capable of reproducing the combined effects seen for all statistics looked at in this study (such as the non-linear total matter power spectrum in Fig. \ref{fig:Pk_HYDRO}, the HMF in Fig. \ref{fig:HMF_HYDRO}, and the two-point autocorrelation function in Fig. \ref{fig:2PCF_HYDRO}) to typically a few percent accuracy. 
    \item As an aside, this study also investigated how well some current approximate methods calibrated on standard $\Lambda$CDM cosmologies can predict some of the results found in this study (see Appendix \ref{appendix: Approx_Methods}). These include a prescription for the HMF produced by \cite{Tinker2008}, as well as a prescription for the redshift evolution of dark matter halo masses, and the c-M relation, produced by \cite{Correa2015c}. We find that for the \cite{Tinker2008} HMF, this is able to reproduce the effects seen due to running out to $z = 2$ very well (Fig. \ref{fig:Tinker_HMF}). We also find that qualitatively, the method developed by \cite{Correa2015c}, is able to reproduce both the evolution of halo mass across the different simulations in this study (Fig. \ref{fig:COMMAH_Mz}); as well as the effect a running scalar spectral index has on the c-M relation of haloes at $z$ = 0 (Fig. \ref{fig:COMMAH_c-M}). 
\end{itemize}
\bigskip
This work has shown that a cosmology with a running scalar spectral index, consistent with the observational constraints of the \cite{PlanckXIII} cosmological parameter results, can have relatively large and measureable effects on LSS. However, it is worth noting that, by itself, running is unable to reconcile the current tension involving $S_8$.  For example, all of the $Planck$ CMB-based runs we examined here have $S_8 > 0.85$ (see Table \ref{table:cosmoparams}), whereas many current LSS tests prefer $S_8$ in the range $[0.75,0.8]$ (see \citealt{McCarthy2018} and references therein).  However, we remind the reader that taking account of the internal tensions in the CMB data, by allowing $A_{\textrm{lens}}$ to float, can significantly reduce the tension between the CMB and LSS (see Appendix \ref{appendix:sim_setup}) and remove it altogether when the neutrino mass is also allowed to float \citep{McCarthy2018}.

Nevertheless, if running is present at the levels suggested by current CMB and Lyman-$\alpha$ forest data, it should be within the detectability range of upcoming surveys such as $LSST$\footnote{\href{https://www.lsst.org/}{https://www.lsst.org/}}, $EUCLID$\footnote{\href{https://www.euclid-ec.org/}{https://www.euclid-ec.org/}}, and $CMB$-$S4$\footnote{\href{https://cmb-s4.org/}{https://cmb-s4.org/}}, and allow tangible constraints to be placed on fundamental, early-Universe physics.

Finally, this study has demonstrated that the effects of a running scalar spectral index are largest at low masses which sample far from the pivot scale.
This is expected as the pivot scale is where all of the power spectra are normalised, and thus forced to have similar values (with the only differences here on these scales arising due to slight differences in $A_s$). 
As mentioned, this scale corresponds to $k_0 =$ 0.05 Mpc$^{-1}$ for $Planck$ CMB measurements, corresponding roughly to cluster scales which \texttt{BAHAMAS} is designed to sample. Therefore, in a future follow-up study we will explore this larger effect on lower masses in more detail, looking at the impact of running on ``near-field" (small-scale) cosmology.

\section*{Acknowledgements}
The authors thank the anonymous referee for a constructive report.
SGS acknowledges an STFC doctoral studentship. This project has received funding from the European Research Council (ERC) under the European Union's Horizon 2020 research and innovation programme (grant agreement No 769130).
This work used the DiRAC@Durham facility managed by the Institute for Computational Cosmology on behalf of the STFC DiRAC HPC Facility. The equipment was funded by BEIS capital funding via STFC capital grants ST/P002293/1, ST/R002371/1 and ST/S002502/1, Durham University and STFC operations grant ST/R000832/1. DiRAC is part of the National e-Infrastructure.




\bibliographystyle{mnras}
\bibliography{Paper_bibliography} 



\appendix
\section{Supplementary}
\label{appendix:sim_setup}
\FloatBarrier

Here we provide the particle mass for each of the 5 simulations, note, the slight differences in the particle masses between the different simulations is due to their slightly different values for $\Omega_{\textrm{m}}$ and $H_0$.
\begin{table}

\caption{(1) Simulation label, (2) the dark matter particle mass in the full hydro simulations, (3) the dark matter particle mass in the dark matter only simulations and (4) the initial baryonic particle mass in the hydro simulations.}
\label{table:part_masses}
\begin{tabular}{cccc}
\hline
(1) & (2) & (3) & (4) \\ \hline
Label & $M_{\textrm{DM,HYDRO}}$ & $M_{\textrm{DM,DMONLY}}$ & $M_{\textrm{bar,init}}$ \\
& (10$^{9}$ M$_{\odot}$ $h^{-1}$) & (10$^{9}$ M$_{\odot}$ $h^{-1}$) & (10$^{8}$ M$_{\odot}$ $h^{-1}$) \\ \hline
-2$\sigma$ & 4.37 & 5.19 & 8.2 \\
-1$\sigma$ & 4.31 & 5.12 & 8.11 \\
$Planck$ ML & 4.32 & 5.15 & 8.25 \\
no-running & 4.36 & 5.17 & 8.11 \\
+1$\sigma$ & 4.44 & 5.26 & 8.13 \\ \hline
\end{tabular}
\end{table}

\begin{figure}
    \centering
    \includegraphics{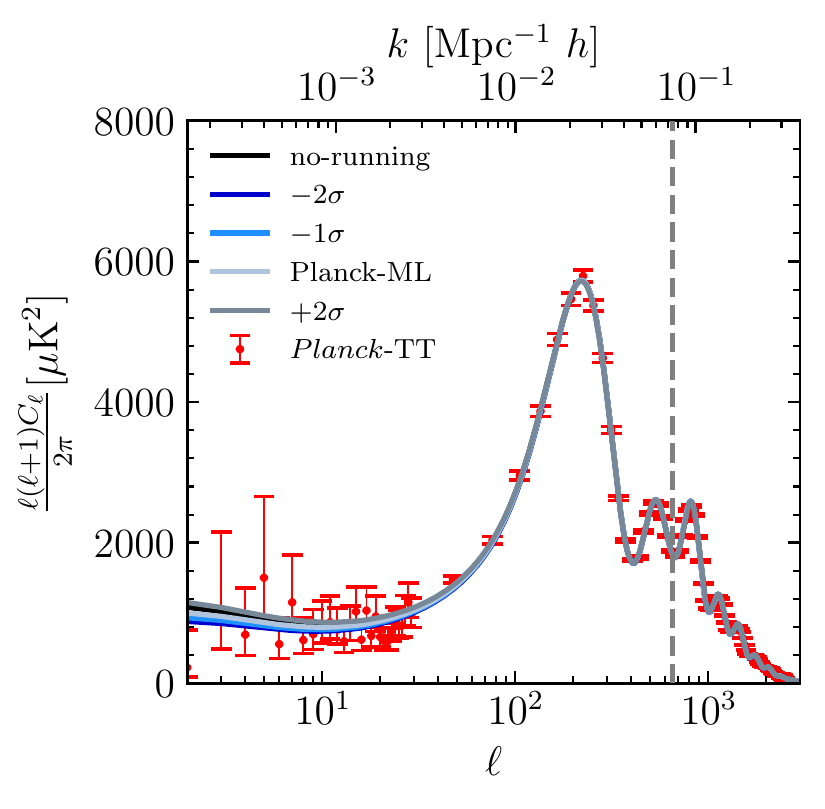}
    \vspace{-0.5cm}
    \caption{The angular power spectrum of temperature anisotropies of the CMB. Data points here are the measurements made by the $Planck$ satellite, and the solid lines represent the theoretical prediction for the different cosmological models. All models are virtually indistinguishable at high-multipoles, with the results at low-multipoles still within observational errors. The corresponding $k-$scales which the $Planck$ observations cover are indicated at the top of the plot. The $Planck$ pivot scale ($k = 0.05 Mpc^{-1}$) is indicated by the vertical grey dashed line.}
    \label{fig:ang_power_spec}
\end{figure}

As mentioned in Section \ref{ParamSelection}, as a test of our cosmological parameter selection we examined the predicted CMB angular power spectra measurements of $Planck$. The result of this can be seen in Fig. \ref{fig:ang_power_spec} and demonstrate that the adopted cosmologies are indistinguishable with regards to current $Planck$ measurements\footnote{\href{https://pla.esac.esa.int/}{https://pla.esac.esa.int/}} for high-multipoles ($\ell \gtrsim 100$), and consistent with $Planck$ measurements at low-multipoles ($\ell \lesssim 100$).

\begin{figure}
    \centering
    \includegraphics[width=\columnwidth]{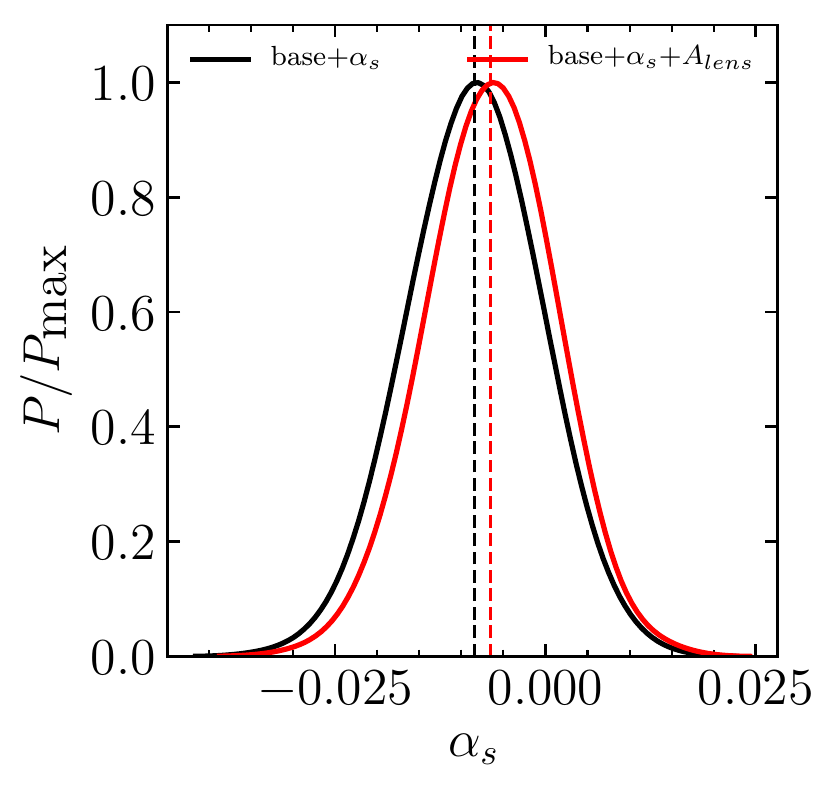}
    \vspace{-0.5cm}
    \caption{Posterior distribution for the running of the spectral index parameter when included in the $Planck$ MCMC analysis, along with $A_{\textrm{lens}}$. Here, the black curve corresponds to constraints on $\alpha_s$ using the $Planck$ 2015 $TT+lowTEB$ dataset, while the red curve shows the same only now $A_{\textrm{lens}}$ is also allowed to vary. The vertical dashed lines correspond to the maximum likelihood value for each distribution. This shows how allowing $A_{\textrm{lens}}$ to also vary in the cosmological model does not affect the constraints on $\alpha_s$ much.}
    \label{fig:1d_Alens}
\end{figure}

Lastly here, we present results from allowing both $A_{\textrm{lens}}$ and $\alpha_s$ to be free.
As mentioned in the main text, for a standard $\Lambda$CDM cosmology, $A_{\textrm{lens}}$ assumes the value of 1. 
However, when also treated as a free parameter in the standard model there is a mild preference for a value greater than unity \citep{PlanckXIII}, which has a knock on effect on other cosmological parameters, such as $H_0$, $\sigma_8$, and $\Omega_{\textrm{m}}$. 
Therefore, we test whether the mild preference for a non-zero negative running which also exists in the $Planck$ 2015 results is mitigated if you no longer fix $A_{\textrm{lens}}$ and $\alpha_s$ to their assumed values of: 1, and 0 respectively.
To do this, we create a new set of chains using the publicly available $Planck$ 2015 likelihood function. We give these parameters the following flat prior ranges:
\begin{center}
    $$A_{\textrm{lens}} \in [0.0, 10.0]$$
    $$\alpha_s \in [-1.0, 1.0]$$
\end{center}

We then select values for $\alpha_s$ and the other important cosmological parameters, including $A_{\textrm{lens}}$ this time, in the same way as that described in Section \ref{sec:Sims}.
The results on the 1D density distribution for $\alpha_s$ are shown in Fig. \ref{fig:1d_Alens}, with the effects on other key parameters for LSS shown in Table \ref{table:params_Alens}.
Looking at Fig. \ref{fig:1d_Alens}, it can be seen that when allowing $A_{\textrm{lens}}$ to also be free, the mild preference for a negative running value is maintained, although the maximum likelihood value is shifted slightly closer to 0. 

As allowing $A_{\textrm{lens}}$ to be free does not drastically impact the resultant values for $\alpha_s$ one would simulate, we focus this study on the one parameter extension to $\Lambda$CDM where we only allow $\alpha_s$ to vary.

\begin{table}
\label{table:params_Alens}
\caption{Cosmological parameter values obtained when treating both $A_{\textrm{lens}}$ and $\alpha_s$ as free parameters. All symbols have their previous meaning, with definitions for all excluding (3) $A_{\textrm{lens}}$ (the lensing amplitude of the CMB TT angular spectrum) present in Table \ref{table:cosmoparams}.}
\begin{adjustbox}{width=\columnwidth,center}
\begin{tabular}{cccccc}
\hline
(1)            & (2)        & (3)        & (4)        & (5)        & (6)     \\ \hline
$\sigma$       & $\alpha_s$ & $A_{\textrm{lens}}$ & $\Omega_{\textrm{m}}$ & $\sigma_8$ & $S_8$   \\ \hline
-2             & -0.02232   & 1.19141    & 0.29153    & 0.82160    & 0.80991 \\
-1             & -0.01442   & 1.19927    & 0.29681    & 0.81539    & 0.81104 \\
Max Likelihood & -0.00652   & 1.21847    & 0.29476    & 0.80750    & 0.80042 \\
+1             & 0.00138    & 1.22956    & 0.29371    & 0.80025    & 0.79182 \\
+2             & 0.00928    & 1.24425    & 0.28979    & 0.79459    & 0.78094 \\ \hline
\end{tabular}
\end{adjustbox}
\end{table}

\begin{figure*}
    \centering
    \includegraphics{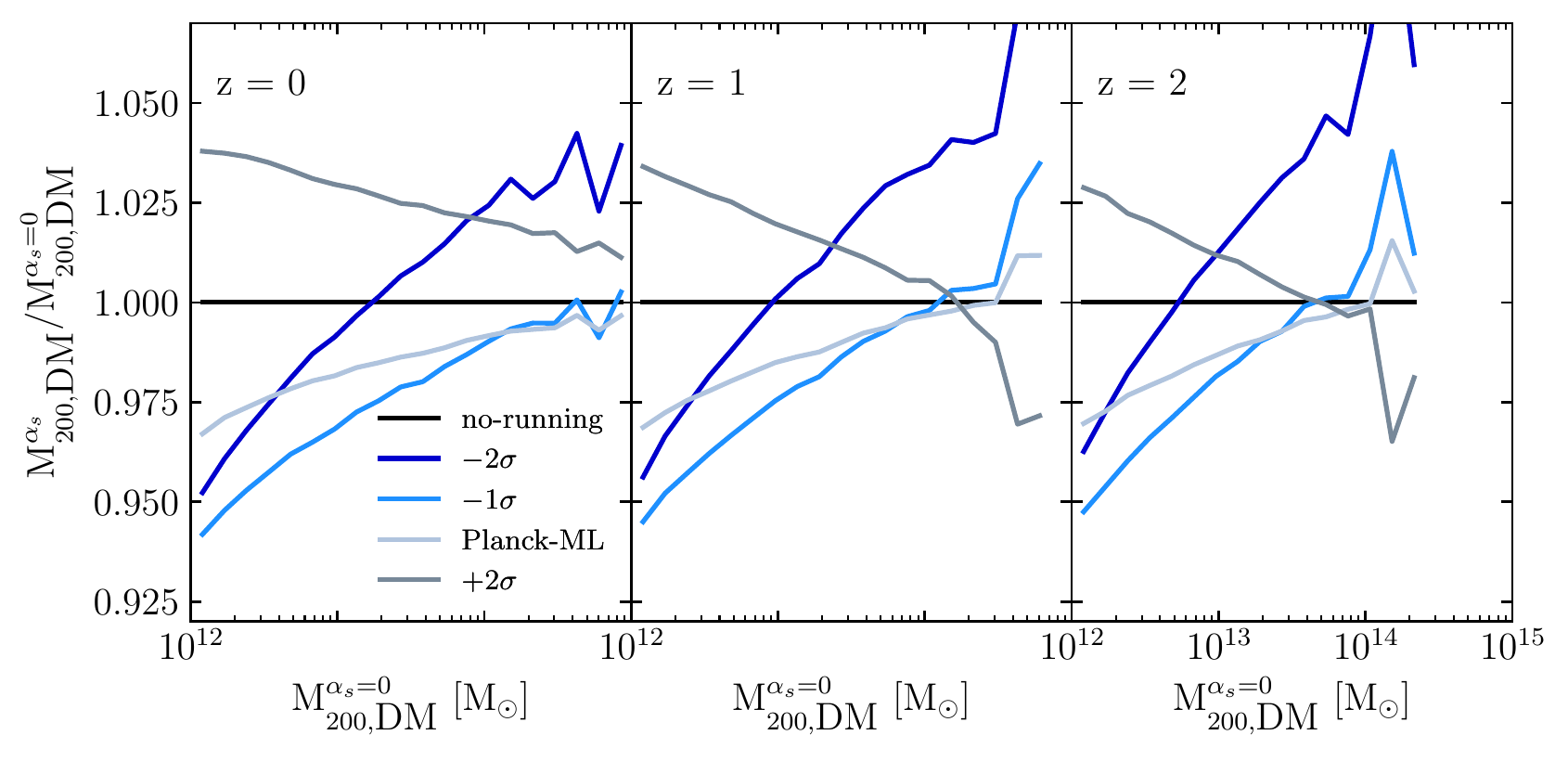}
    \vspace{-0.5cm}
    \caption{The fractional change in halo mass for a matched set of haloes across the 5 different cosmologies for 3 separate redshifts (indicated in the top-left corner of each panel). This shows that a halo in a negative running cosmology is more massive than it would be in a no-running cosmology, with this effect amplified with redshift. Conversely a halo in a positive running cosmology is more massive at the present day, but less massive as you go back in redshift.}
    \label{fig:frac_mass_change_z_evolution}
\end{figure*}
Finally in this section we show in Fig. \ref{fig:frac_mass_change_z_evolution} the redshift evolution of the fractional change in halo mass. This helps explain the results which were seen in the redshift evolution of the HMF (Fig. \ref{fig:HMF_DM}).
\FloatBarrier

\section{Approximate Methods}
\label{appendix: Approx_Methods}

Although a systematic exploration of the effects a running scalar spectral index on LSS in the Universe using large cosmological hydrodynamic simulations has not been conducted, research has went into the effects of running using semi-analytic models (e.g. \cite{Fedeli2010}).
Here, we explore how well these approximate methods work in reproducing the HMF, and the redshift evolution of dark matter haloes. 
\FloatBarrier
\subsection{Halo mass function}
\begin{figure}
    \centering
    \includegraphics[width=\columnwidth]{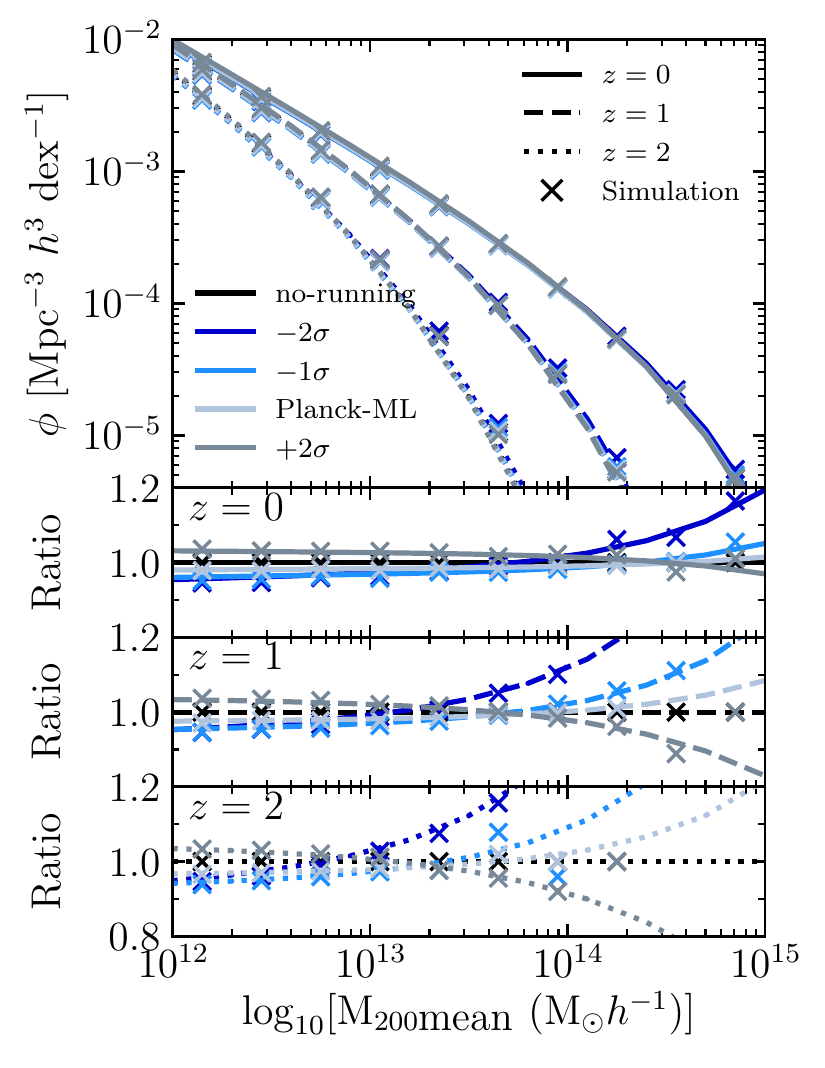}
    \vspace{-0.5cm}
    \caption{Top: the halo mass function (HMF) for the 5 separate cosmologies looked at in this study. 
    Here, the lines represent the halo mass function calculated according to the \citet{Tinker2008} prescription, with the linestyles indicating the different redshifts for which the HMF was computed. Also shown in this plot is the HMF measured from the simulations themselves, indicated by the crosses.
    Note that here, unlike the HMF shown in Fig. \ref{fig:HMF_DM}, the HMF is plotted in terms of $M_{200,\textrm{mean}}$ and not $M_{200,\textrm{crit}}$. 
    The bottom panel shows the HMF normalised with respect to the $\alpha_s = 0$ case, for each respective method, at each redshift.
    It can be seen that the \citet{Tinker2008} HMF reproduces the effects due to a running scalar spectral index remarkably well.}
    \label{fig:Tinker_HMF}
\end{figure}

First, we look at how well the prescription for the HMF provided in \cite{Tinker2008} (T08) is able to describe the HMF that we measure for the 5 separate cosmologies from the simulations. 
The resultant plot can be seen in Fig. \ref{fig:Tinker_HMF}, which shows the HMF at three separate redshifts generated using the prescription of T08, represented with the different curves.
Also shown is the actual HMF measured from the simulations. 
Note that here, we plot the HMF as a function of $M_{200,mean}$ instead of $M_{200,\textrm{crit}}$ as was done previously.
Here $M_{200,mean}$ describes the mass enclosed in a spherical overdensity, whose radius contains an overdensity equal to 200 $\times$ the mean density of the Universe ($\rho_{\textrm{c}} \Omega_{\textrm{m}}$). 
The bottom panel of this plot shows the HMFs normalised with respect to the $\alpha_s = 0$ model, for each case, and at each redshift. 
This plot shows that the T08 HMF can reproduce the measured HMF remarkably well at $z = 0, 1, 2$. 
There is a hint from the top plot that there is a slight under prediction of the number of high-mass haloes at $z = 2$, but the relative effect shown in the bottom panel is almost perfectly reproduced.

It is important to highlight that, the above T08 mass function was computed by using the linear matter power spectrum output using \texttt{CAMB} to compute the mass variance $\sigma(M)$. 
Off the shelf packages such as \texttt{HMFCALC}\footnote{\href{http://hmf.icrar.org/}{http://hmf.icrar.org/}} are not able to reproduce the observed HMF, due to the fact that they are hardwired to use the transfer function when computing the power spectrum at a certain redshift, with their primordial matter power spectrum assuming a constant power law. 

\subsection{Mass accretion histories}
We now look at the effects a running scalar spectral index has on the the evolution of dark matter haloes. We do this using the method developed by \cite{Correa2015a} \citep[see also][]{Correa2015b, Correa2015c} to reproduce the mass-accretion histories of haloes, along with the concentration-mass relation.

We used a slightly modified version of the python package developed by \cite{Correa2015a}: \texttt{COMMAH}\footnote{\href{https://bitbucket.org/astroduff/commah/src/master/}{https://bitbucket.org/astroduff/commah/src/master/}}. We modify it to account for a running scalar spectral index in its calculation of the $z = 0$ linear matter power spectrum. Following \cite{Correa2015c}, we also calculate a value for $A_{\textrm{cosmology}}$, for each cosmology in order to compute the c-M relation.

\begin{figure}
    \centering
    \includegraphics[width=\columnwidth]{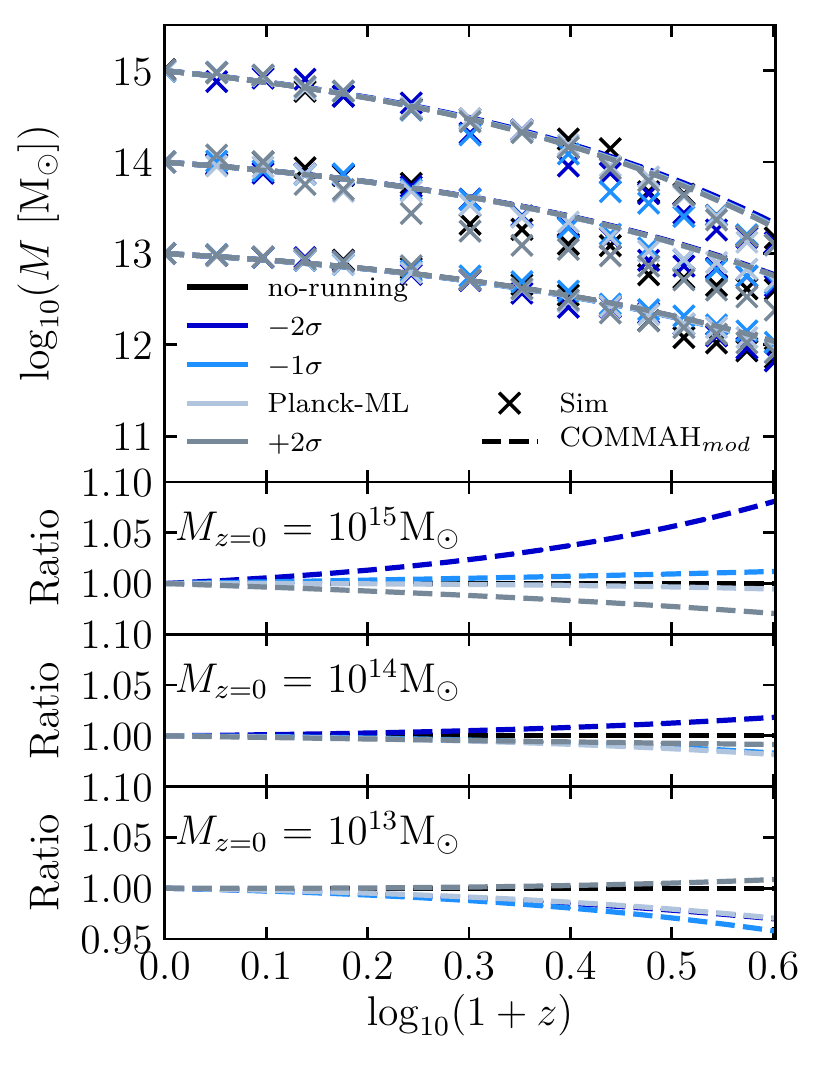}
    \vspace{-0.5cm}
    \caption{Evolution of halo mass, with redshift. Top: the dashed lines here indicate the predicted $M(z)$. The crosses show the results obtained directly from the simulations. The general trend of these results are in qualitative agreement. Bottom: the ratio of the \texttt{COMMAH} predicted $M(z)$, with respect to the no-running cosmology, for three separate halo masses, indicated in each panel. It appears to show again, that a running scalar spectral index has a mass dependent effect on haloes. With larger-mass haloes being more massive at early times in a negative running cosmology, and lower-mass haloes, less massive. This effect is reversed somewhat for the positive running cosmology.}
    \label{fig:COMMAH_Mz}
\end{figure}

The redshift evolution of halo mass can be seen in Fig. \ref{fig:COMMAH_Mz}. 
Here the dashed curves show the result obtained using the modified \texttt{COMMAH} code, with crosses representing the evolution of halo mass with redshift computed from the actual simulations.
To compute this, we choose haloes in a narrow mass bin centered on the desired redshift 0 mass.
We then find that halo's progenitor at each snapshot up to $z = 3$. 
This is done by matching the 50 most bound particles from the $z = 0$ halo, and selecting the halo at each redshift which contains the most of these particles.
It can be seen that there is a qualitatively good agreement between the simulation's result and that predicted using \texttt{COMMAH}. 
The bottom panels on this plot show the predicted redshift evolution of three haloes of different mass, normalised with respect to the no-running evolution for that mass. 
It can be seen that the introduction of a running spectral index affects how these haloes evolve. For example, the $-2\sigma$ cosmology predicts a $10^{15}M_{\odot}$ halo to be around 10$\%$ more massive at $z = 3$, compared with a no-running cosmology.
Whereas, for that same cosmology, a $10^{13} M_{\odot}$ is near 5$\%$ under-massive compared with a no-running cosmology.
This general trend agrees with the mass-dependent effects seen in this study for the HMF.
It also implies that the MAH of these objects are going to be different, which we examine next. 

\begin{figure}
    \centering
    \includegraphics[width=\columnwidth]{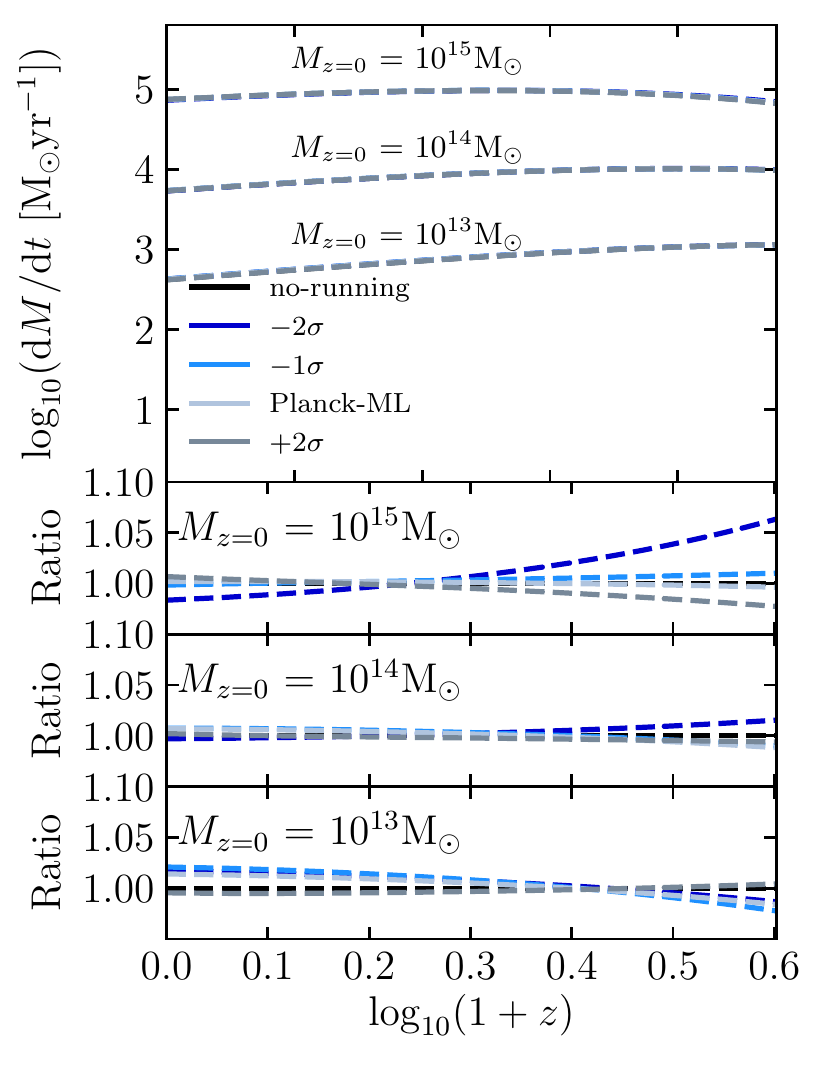}
    \vspace{-0.5cm}
    \caption{Mass accretion histories (MAHs) predicted using the slightly modified version of the \texttt{COMMAH} python package for 3 different halo masses for each cosmology examined here. Top: MAH predicted for 3 different halo masses, indicated above each curve. The bottom panels how the MAH of each halo normalised to its respective no-running counterpart. How a halo forms in a cosmology which has a running scalar spectral index depends on both the sign and magnitude of the running, along with the mass of the object.}
    \label{fig:COMMAH_dMdt}
\end{figure}

The predicted MAH for the same set of halo masses as examined previously is shown in Fig. \ref{fig:COMMAH_dMdt}. 
The bottom panels in this plot show the MAH for the 5 cosmologies, normalised with respect to the result in the no-running cosmology at that mass.
This result helps explain Fig. \ref{fig:COMMAH_Mz}. For example, looking at the $10^{15}M_{\odot}$ halo's MAH, it can be seen that for the $-2\sigma$ cosmology, it accretes a lot of its mass at early times, causing it to be more massive earlier on. 
Conversely, in the positive running cosmology, it has a steady incline in its mass accretion rate, compared with the no-running cosmology, and has a higher accretion rate at present. 
This allows haloes of this mass, in a positive running cosmology, to start with a smaller seed mass, compared to no-running. Similarly, for the $10^{13}M_{\odot}$ halo, for all negative running cosmologies, $M(z)$ is lower than the no-running prediction, but this is made up for, by an increase in the MAH of these haloes as $z$ approaches 0. 

\begin{figure}
    \centering
    \includegraphics[width=\columnwidth]{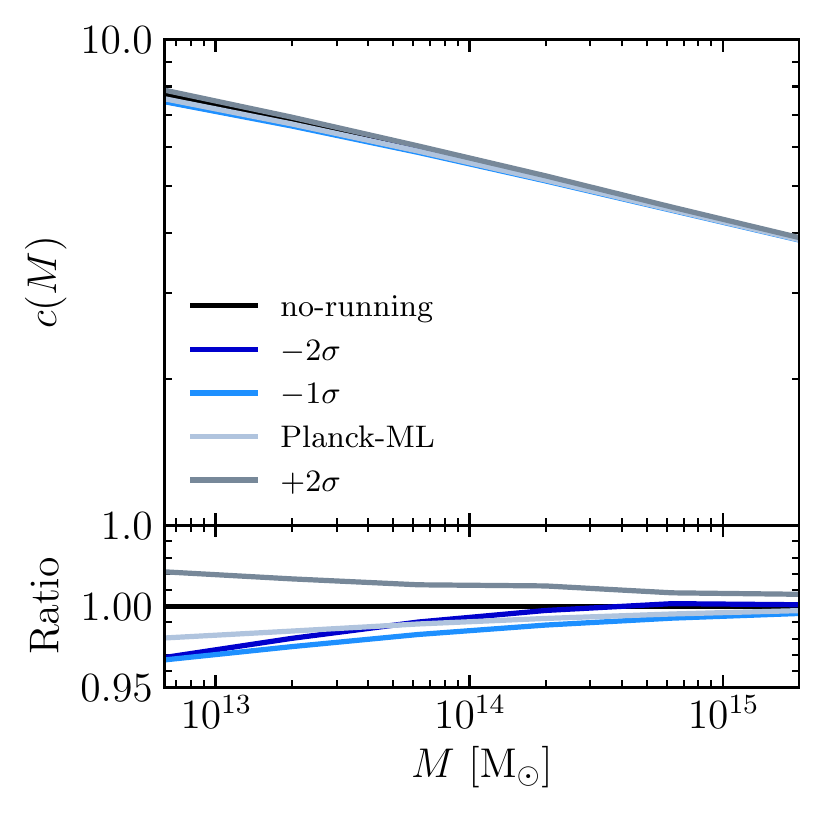}
    \vspace{-0.5cm}
    \caption{The halo concentration-mass relation for the different running cosmologies predicted using the slightly modified \texttt{COMMAH} python package. The bottom panel of this plot shows the result normalised with respect to the no-running cosmology. A negative running cosmology is predicted to have less concentrated lower-mass haloes, whereas a positive running yields higher concentrations. This is in qualitative agreement to the results found in this study (see Fig. \ref{fig:c-M_relation}). Conversely, for larger-mass haloes only a positive running cosmology predicts higher concentrations, whereas it was found in this study that all cosmologies with a non-zero running spectral index had higher concentrations compared to the no-running result.}
    \label{fig:COMMAH_c-M}
\end{figure}

Finally, we show the predicted halo concentration-mass relation for these 5 separate cosmologies in Fig. \ref{fig:COMMAH_c-M}.
The bottom panel shows the halo concentration-mass relation normalised with respect to the no-running cosmology.
Qualitatively, the results are similar to what was found from the simulations. 
For example, a positive running cosmology predicts all haloes in the mass-range examined here to have a higher concentration, compared to the no-running cosmology. Alongside lower-mass haloes ($M \lesssim 2 \times 10^{13}$M$_{\odot}$) having lower concentrations in a negative running cosmology. However, whereas in the simulations all haloes above $M \approx 10^{14}$M$_{\odot}$ were found to have higher concentrations, here they are predicted to have slightly lower concentrations, with the results converging towards (but not exceeding) the no-running prediction at the high-mass end.

\bsp	
\label{lastpage}
\end{document}